\documentclass[useAMS,usegraphicx]{mn2e}

\def\etal{{\it et al.~}}
\def\kms{km s$^{-1}$~}
\def\lsim{\mathrel{\rlap{\lower4pt\hbox{\hskip1pt$\sim$}}    \raise1pt\hbox{$<$}}} 

\title[NGC~5128 Globular Clusters]
{A 2dF Spectroscopic Study of Globular Clusters in NGC~5128: Probing the
Formation History of the Nearest Giant Elliptical}

\author[M. Beasley \etal]{
Michael A. Beasley$^{1}$, 
Terry Bridges$^{2}$, 
Eric Peng$^{3}$, 
William E. Harris$^{4}$, 
\newauthor
 Gretchen L.H. Harris$^{5}$, 
{Duncan A. Forbes$^{6}$
and Glen Mackie$^{6}$}\\
\thanks{email:beasley@iac.es},
\thanks{tjb@astro.queensu.ca},
\thanks{Eric.Peng@nrc-cnrc.gc.ca},
\thanks{glharris@astro.uwaterloo.ca}, 
\thanks{harris@physics.mcmaster.ca},
\thanks{dforbes@astro.swin.edu.au}, 
\thanks{gmackie@astro.swin.edu.au}
\\
$^1$Instituto de Astrof\'isica de Canarias, 
Via Lactea, E-38200 La Laguna, Tenerife, Spain\\
$^2$Department of Physics, Queen's University, Kingston, 
ON K7L 3N6, Canada\\
$^3$Herzberg Institute of Astrophysics, National Research
Council of Canada, 5071 West Saanich Road, Victoria, 
BC V9E 2E7, Canada\\
$^4$Department of Physics and Astronomy, McMaster University, 
Hamilton, ON L8S 4M1\\
$^5$Department of Physics and Astronomy, University of Waterloo, Waterloo, 
ON N2L 3G1, Canada\\
$^6$Centre for Astrophysics and Supercomputing, Swinburne
University of Technology, Hawthorn, VIC 3122, Australia\\
}

\date{\today}

\begin{document}
\maketitle

\label{firstpage}

\begin{abstract}
We have performed a spectroscopic study of globular clusters (GCs)
in the nearest giant elliptical NGC~5128 using the 2dF facility
at the Anglo-Australian telescope. We obtained integrated 
optical spectra for a total of 254 GCs, 79 of which are 
newly confirmed on the basis of their radial velocities and spectra.
In addition, we obtained an integrated spectrum of the galaxy starlight 
along the southern major axis.
We derive an empirical metallicity distribution function (MDF) for 207 GCs 
($\sim14\%$ of  the estimated total GC system) based upon Milky Way GCs.
This MDF is multimodal at high statistical significance with peaks 
at [Z/H]$\sim-1.3$ and $-0.5$. 
A comparison between the GC MDF and that of the stellar halo
at 20 kpc ($\sim 4$R$_e$) reveals close coincidence at the metal-rich 
ends of the distributions. However, an inner 8 kpc stellar MDF shows
a clear excess of metal-rich stars when compared to the GCs.
We compare a higher S/N subsample (147 GCs) with two stellar population models 
which include non-solar abundance ratio corrections.
The vast majority of our sample ($\sim90\%$) appears old, with ages similar to the 
Milky Way GC system. 
There is evidence for a population of intermediate-age ($\sim4-8$ Gy) GCs
($\leq15\%$ of the sample) which are on average more metal-rich than the
old GCs.  We also identify at least one younger cluster ($\sim1-2$ Gy)
in the central regions of the galaxy. Our observations are consistent 
with a picture where NGC 5128 has undergone at least two mergers and/or 
interactions involving star formation and limited GC formation since $z=1$, 
however the effect of non-canonical 
hot stellar populations on the integrated spectra of GCs remains an 
outstanding uncertainty in our GC age estimates.
\end{abstract}
%
\section{Introduction}
\label{Introduction}

One of the principal tenets of extragalactic GC research is that 
``GCs trace the major episodes of star formation in their host galaxies''.
Supporting evidence for this view  includes the fact that 
the mean colours (a surrogate for metallicities at old ages) of blue and red 
GC subpopulations in  galaxies are correlated with the host galaxy luminosity
(e.g., Forbes, Brodie \& Grillmair 1997; Larsen \etal2001; 
Strader, Brodie \& Forbes 2004; Peng \etal2006; Brodie \& Strader 2006).

Further evidence that GCs may be used as proxies for star formation in 
galaxies 
comes from recent spectroscopic observations of the 
GC systems of early-type galaxies (Es).  Strader \etal(2005) find little 
evidence of young GCs from a small sample of  massive Es, in agreement
with previous studies using either spectroscopy
(Cohen, Blakeslee \& Rhysov 1998; Beasley \etal2000)
or the measured luminosity function with the assumption of a 
universal mass function for all GCs (Puzia \etal1999;  Jord{\'a}n \etal2002).
By contrast, Puzia \etal(2005) find evidence to suggest that up to 30\% of 
GCs in less massive Es may have formed later than $z=1$.
Whether these younger GCs formed in dissipational mergers or by some other 
means is presently unclear. The finding of young GCs in lower-mass 
ellipticals has found some support in results from IR photometry 
(Hempel \etal2007), which in conjunction with 
optical colours promises to break the age-metallicity degeneracy 
(e.g., Mould \& Aaronson 1980). 
However, this technique is proving somewhat controversial 
(Larsen \etal2005; Kundu \etal2005), and the reliability of this 
approach needs to be fully tested both observationally and theoretically
(e.g., Salaris \& Cassisi 2007; Lee \etal2006).

Notwithstanding ongoing argument about the use of IR photometry in age 
estimations, spectroscopy is still the method of 
choice for deriving ages and metallicities, and indeed is the 
only way to estimate abundance ratios and obtain kinematic information. 
The current picture from the spectroscopic studies is, qualitatively 
at least, in agreement with the ``downsizing'' phenomenon. Massive 
ellipticals are old, passively evolving systems while their less massive 
counterparts continue to merge and make stars and GCs at later epochs.
Unfortunately, spectroscopic sample sizes of GC systems remain small.
Most of the nearest massive Es are at Virgo distances which restricts samples
to the brightest few percent of clusters on 8-m class telescopes. 
This not only places statistical limitations on the interpretation of 
data, but also raises the question of whether the most massive globular 
clusters are representative of the system at all (e.g., Harris \etal2006a).

An obvious target with which to begin to remedy this problem of 
sample sizes is NGC 5128, the nearest giant elliptical galaxy. 
At a distance of $\sim4$ Mpc, over half of the GC system is 
amenable to spectroscopic observations (with the brightest third or so accessible 
to 4-m class telescopes). Moreover, NGC 5128 shows extensive evidence for 
one or more recent interactions, including shells (Malin 1978), a central, highly
inclined gas and dust disk (e.g., Dufour \etal1979), and a stream of young
stars in its halo (Peng \etal2002). Indeed, in view of all this recent 
activity in the galaxy, if the association between the GCs and 
field stars in ellipticals is to be understood, then NGC 5128 is 
an excellent place to start. 

Peng, Ford \& Freeman (2004) (hereafter PFF04) were the first to look in any detail at the 
spectroscopic ages and metallicities of GCs in NGC 5128, building on previous 
work by van den Bergh, Hesser \& Harris (1981), Hesser, Harris \& Harris (1986), 
Sharples (1988) and Held \etal(2002). 
Based upon spectra of 23 GCs with sufficient S/N, PFF04 found that
the NGC 5128 GCs displayed a wide range of metallicities, and that the 
metal-rich GCs ([Fe/H]$>-1$) had a spread in ages, with a mean 
age of $\sim5$ Gy.
Based in part on the results for the GCs, PFF04 
suggested that the NGC 5128 galaxy which we see today was assembled through the 
dissipational merger of two unequal-mass disks at $z<1$.
Bekki \& Peng (2006) have modelled the same merger process
to explain the kinematics and spatial distribution of
its planetary nebula population.

In this paper we present the first results of an ongoing spectroscopic survey of 
the GC system of the nearby E/S0 NGC~5128, examining the ages, metallicities 
and abundance ratios of GC system in statistically significant numbers. 
The plan of this paper is as follows:
In Section~\ref{TheData} we describe our candidate GC sample, observations and 
data reduction.
In Section~\ref{Analysis}
we derive metallicity, age and abundance ratio estimates for the GCs and 
discuss properties of the NGC 5128 GC system.
In Section~\ref{Summary}, we present a summary of our results 
and a discussion.
A detailed analysis of the kinematics of the GC system is presented
in Woodley \etal(2007), and a discussion of the colour-metallicity relations
will be forthcoming (Peng \etal2007 in preparation). 

Throughout the following, 
we adopt a distance of 3.8~Mpc to NGC~5128 (Rejkuba 2004).
At this distance, 1\arcsec $\sim$ 20pc. 

\section{The Data}
\label{TheData}

The principal goals of our spectroscopic survey were twofold; to 
obtain optical spectra of GC candidates in NGC~5128 in order to increase 
the number of known {\it bona fide} GCs, and to obtain spectra for previously 
known, and newly identified clusters of sufficient quality for a metallicity, 
age and kinematic analysis of the GC system. We describe our observational 
approach in this section.

\subsection{Observations and Data Reduction}
\label{Observations}

The wide angular extent of the NGC~5128 GC system on the sky and the 
large number of candidate GCs in our sample required high multiplex 
over a wide-field. 
These requirements made the 2-degree field (2dF) instrument on the 
Anglo-Australian Telescope a logical choice.  The 2dF permits up to 
400 spectra to be obtained simultaneously over a 2$\degr$.1 diameter 
field of view (see Lewis \etal2002). The fibres are fed to two 
spectrographs with 200 fibres per spectrograph. Each fibre has a 
$\sim2$~arcsec diameter, matching well the NGC~5128 GCs which are 
unresolved at the median seeing of Siding Spring ($\sim$1.3~arcsec). 
Our observational set-up is given in Table~\ref{tab:setup}. 
We observed previously known GCs taken from Peng \etal(2004b), as well
as new GC candidates.  These candidates were selected from the UBVRI
photometry presented in Peng \etal(2004b, see their section 4 for a detailed
description of the selection), and augmented with the wide-field
Washington system photometry of Harris \etal(2004) to yield a master
list of $\sim 1000$ candidates to $V\sim 20.5$.  We identified GC candidates
from both sets of photometry using a combination of colour and
morphological cuts that selected objects appearing slightly extended
and having the colours of GCs.  We also matched our catalogues
with the X-ray point source detections of Kraft \etal(2001) and
assigned a high priority to any optical counterparts since low-mass X-ray
binaries are known to be associated with GCs (Clark 1975).

Spectra for GC candidates were obtained between the nights of May 6 and 
May 11 in 2003. Two and a half nights were lost to bad weather and the 
remaining nights were clear. To maximise efficiency, our observing 
strategy was to first observe three ``shallow'' (3 hr) fields which 
encompassed all cluster candidates with no spectroscopic confirmation. 
These data were reduced on-the-fly at the telescope using the 2dF pipeline 
software {\sc 2dfdr} (Taylor \etal1996; Bailey \etal2002). Radial velocities 
were measured using {\sc 2dfgrs}, which implements a cross-correlation 
approach against empirical template spectra (Colless \etal2001).  
Selecting by velocity (Section~\ref{RadialVelocities}), an essentially clean 
sample of GCs was then constructed with which we made two plate 
configurations for deeper (8 hr) spectroscopy. Some 40 fibres were 
allocated to sample the sky 
uniformly across the field for sky subtraction. A further 10 fibres were 
positioned $1\arcmin-10\arcmin$ ($\sim1.2-12$ kpc) south of the galaxy centre along 
the major axis to sample the integrated galaxy light, but at the 
same time avoiding the prominent dust/gas lane in the galaxy.

The 10 galaxy fibre allocations were crucial for accurate 
removal of the galaxy background from the GC spectra, and for synthesizing 
an integrated spectrum of the galaxy for subsequent analysis. 
The flux-weighted combined galaxy spectrum corresponded to a position
of approximately 2.3 arcmin south along the major axis. 
Adopting a V-band effective radius of 5.0 arcmin for NGC 5128 (Dufour \etal1979), 
this corresponds to approximately an r$_e/2$ aperture. 
In addition, spectra common between spectrographs 1 and 2 
(from different plate configurations) were obtained for 11 previously identified 
GCs spanning a range of magnitudes. This allowed us to estimate the 
internal uncertainties of our radial velocity and linestrength measurements.

The sky and galaxy fibre allocations were used to remove the contribution
from the sky and galaxy background from our GC observations. Beyond approximately 
10 arcmin, the flux contribution from the spheroid 
was minimal, and an averaged sky spectrum created from the combined sky fibres 
was subtracted from the spectra of GCs beyond this radius. For GCs projected 
interior to this radius, the galaxy spectrum obtained along the major axis was 
scaled to the position of each GC (by fitting a smooth curve through a plot of 
the mean flux down each galaxy fibre versus radius) and subtracted 
from each GC (similarly, see Woodley, Harris \& Harris 2005 and PFF04).

\subsection{Radial Velocities}
\label{RadialVelocities}

\begin{figure}
\centerline{\includegraphics[width=8.0cm]{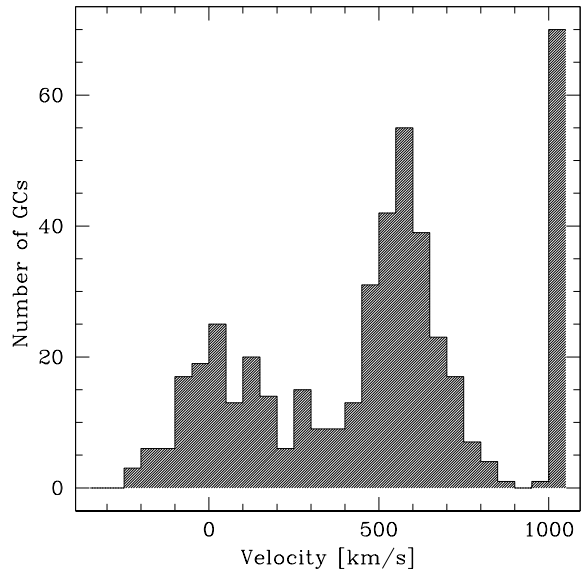} }
\caption{Histogram of 454 GC candidates for which velocities could be 
measured. The GC system of NGC~5128 is clearly visible with a mean velocity
of $\sim550$ \kms. The distribution of objects with a mean velocity of 
$\sim50$ \kms are foreground Galactic stars. The pile-up at  
1200 \kms are background galaxies and is artificial.
\label{fig:VelocityDistribution}}
\end{figure}

Preliminary radial velocities for all candidate GCs were obtained 
with {\sc 2dfdr} as outlined in Section~\ref{Observations}. 
We found no systematic differences in velocities of common 
objects between the spectrographs, however, during the observations, 
differences in the velocity 
zeropoints of order $\sim30$ \kms were observed between standard stars 
with different plate configurations.
Therefore, post-observing, GC velocities were re-determined with {\sc fxcor}
in {\sc iraf}\footnote{IRAF is distributed by the National Optical Astronomy 
Observatories, which are operated by the Association of Universities for 
Research in Astronomy, Inc., under cooperative agreement with the National 
Science Foundation.} using 12 F--K stellar templates which doubled 
as Lick standard stars.  
The velocity zeropoints of these templates were checked by cross-correlating 
them with synthetic spectra (Vazdekis \etal2007, in preparation). 
Velocities were considered secure if they had a clearly defined cross-correlation 
peak, and a Tonry \& Davis (1979) $r$ value $>3$. 
The velocity from the best-matching template was then adopted as the final velocity.

We noted during the analysis stage that there existed resolution variations
in these data of up to $\sim20\%$ as a function 2dF plate position. 
We verified that no systematic uncertainties were introduced in our
measured radial velocities by smoothing all the spectra to a common resolution
and re-measuring velocities. The only effect of the varying resolution was to 
increase our velocity uncertainties as a function of 2dF plate position.
It should be noted though, that for index measurements the spectra were
smoothed to a common resolution (Section~\ref{LickIndices}) significantly 
lower than the instrumental resolution of our setup.

The velocity distribution of all the candidate clusters is shown in 
Figure~\ref{fig:VelocityDistribution}. 
As can be seen, the velocity distribution of NGC~5128 GCs is 
well separated from background galaxies and reasonably well 
separated from Milky Way foreground stars.
Following Peng \etal(2004b), and informed by Fig~\ref{fig:VelocityDistribution}, 
we classify as GCs all candidates in the velocity interval of 250--1000 \kms. 
At the lower bound, 250 \kms misses very few GCs.  For the field stars in the NGC
5128 planetary nebula (PN) sample of Peng, Ford, \& Freeman (2004c),
which is free from contamination, only 1\% of PN velocities are
below this cutoff.
However, Fig~\ref{fig:VelocityDistribution} does suggest that some 
uncertainty in star/GC identification remains for cluster candidates with 
V$\sim$250--350 \kms. Indeed, one of these objects could be clearly identified 
as a foreground star on the basis of its spectrum (object AAT304867 has V=$305\pm55$ \kms, 
but exhibits strong molecular bands in its spectrum characteristic of an M-type star). 
Another (AAT114993) has been identified as a star on the basis of {\it HST}/ACS imaging 
(Harris \etal2006b) despite its having V=$352\pm136$ \kms 
(though note the large velocity uncertainty).

In an attempt to quantify the level of contamination from foreground stars 
in this velocity range we ran KMM (Ashman, Bird \& Zepf 1994) on the combined 
velocity distribution of stars and GCs assuming two homoscedastic 
populations\footnote{Assuming two heteroscedastic populations in KMM
made little difference}. 
KMM gave estimated means of 53 and 564 \kms for the stars and GCs 
respectively, with a 50\% posterior probability that an object belongs 
to either subpopulation (based purely on its radial velocity) occurring at 
a velocity of 290 \kms. At 350 \kms KMM predicts that the posterior 
probability that an object is in fact a GC is $\sim90\%$. This probability 
falls to $\sim20\%$ at 250 \kms. There are 25 GC candidates in the velocity 
range 250--350 \kms, and based on KMM we estimate 
that approximately half of these may be foreground stars. We note that this 
potential contamination from foreground stars represents $<5\%$ of our GC sample.

\begin{figure}
\centerline{\includegraphics[width=8.0cm]{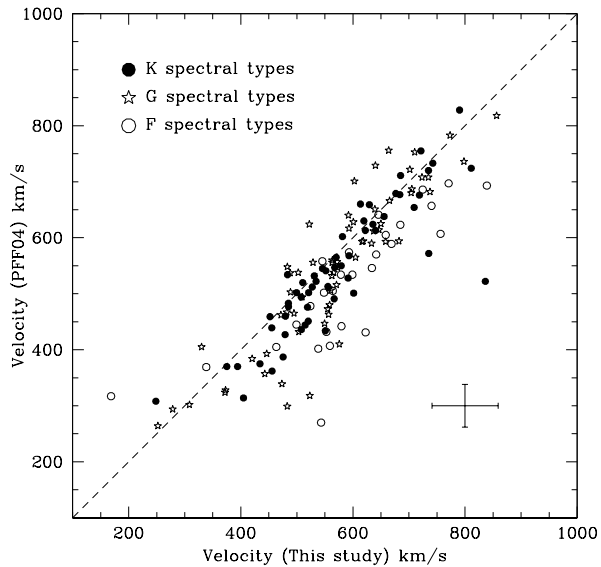} }
\caption{Comparison of 148 velocities in common between this study and 
PFF04. The dashed line is the 1-to-1 relation. Symbols correspond
to the spectral type of the best cross-correlation template.
\label{fig:CompareEric}}
\end{figure}

In Fig~\ref{fig:CompareEric} we compare our velocities with those from the 
study of Peng \etal(2004b), observed with a combination of 2dF and Hydra on the CTIO 4-m.
We find good agreement between the two samples, but note a systematic offset
in the sense of us-Peng \etal = +35 \kms, with a dispersion of 66$\pm5$ \kms.
The dispersion is consistent with our mean velocity uncertainty (59$\pm4$ \kms), 
though the source of the offset is unclear. We have re-checked the zeropoints
of our template stars and heliocentric corrections and find no obvious
systematic error. We separated our velocities by the spectral 
type of the best matching templates (F G \& K types; Fig~\ref{fig:CompareEric}) 
and find the largest discrepancy for F-type GCs (us-Peng \etal = +74 \kms, 
with a dispersion of 77 \kms).
The G \& K templates give offsets of 30 and 23 \kms respectively, with 
identical dispersions of 60 \kms. Since the NGC 5128 GC system clearly spans 
a range of spectral types, it is possible that use of a single template
for cross-correlation in Peng \etal (M31 GC 225-280) may have led to systematic
variations in velocity as a function of spectral type.

\begin{figure}
\centerline{\includegraphics[width=8.0cm]{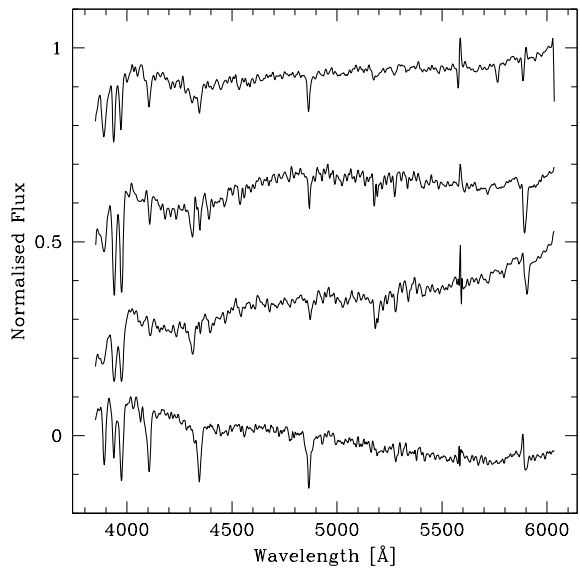} }
\caption{Example 2dF spectra of NGC~5128 GCs with median S/N (40--60 \AA$^{-1}$). 
From top to bottom GCs are AAT109711 (metal-poor, old), 
HGHH-21 (intermediate metallicity, old), K-102 (metal-rich, old) and 
HGHH-279 (intermediate metallicity, young).
Spectra are shown at the Lick/IDS resolution ($\sim8-11$~\AA).
\label{fig:spectra}}
\end{figure}

Of 454 unique spectra for which we could derive reliable velocities, 
we classify 70 as galaxies, 130 as foreground stars and 254 as GCs 
belonging to NGC~5128. Of these 254 GCs, 79 had no previous velocity measurements. 
In Table~\ref{tab:GCs} we list the velocities and 
co-ordinates of all GCs observed. The newly identified GCs are designated
``AAT$\#$'' in Table~\ref{tab:GCs}, previously known GCs retain their
previous designations (see PFF04). 
For completeness, we list the
GC candidates which turned out to be stars or galaxies in 
Tables~\ref{tab:stars} and \ref{tab:galaxies}. 
We show a selection of the 2dF spectra in Figure~\ref{fig:spectra}.

\subsection{Lick Indices}
\label{LickIndices}

In order to compare our spectra to single stellar population models (SSPs), 
we calibrated them to the Lick system (Trager \etal1998) using 
12 Lick standard observed with the same 2dF configuration (Section~\ref{RadialVelocities}). 
The standard stars and the NGC~5128 
GC spectra were smoothed with a wavelength-dependent Gaussian to the resolutions 
given in Worthey \& Ottaviani (1997). The Lick index definitions of 
Worthey \& Ottaviani (1997) and Trager \etal(1998) were measured for 
all spectra, and additive offsets from the Lick standards applied to 
the GC data. These offsets were obtained by comparing our Lick index 
measurements of the standard stars to the Lick/IDS 
measurements\footnote{http://astro.wsu.edu/worthey/html/system.html}. 
Straight additive 
offsets (obtained using a weighted least-squares fit) 
were applied since there was little evidence of non-linearity in 
the index comparisons as a function of metallicity. 
The total uncertainty in our index measurements was
calculated as the quadrature addition of the Poisson uncertainty, the repeat 
measurements uncertainty and the correction to Lick uncertainty. 
For most indices, the correction to the Lick system was by far the dominant uncertainty.
Our Lick index measurements are listed in Table~\ref{tab:LickIndices} and 
index uncertainties are given in Table~\ref{tab:LickErrors}.

\section{Analysis}
\label{Analysis}

\subsection{An empirical metallicity scale}
\label{MetallicityScale}

Before turning to SSPs
to derive population parameters, we wished to derive a purely empirical, 
model-independent metallicity scale for our data in order to rank the 
NGC~5128 GCs by their metal abundance. There is currently no metallicity 
scale which has been universally adopted in extragalactic  
GC research, but those which are used, derived either from integrated 
colours or spectroscopy, are invariably tied to either theoretical isochrones 
or Milky Way GC metallicities. We have chosen the latter route, and as a 
calibration set use Lick index data available for  Milky Way GCs. 
We used GC data for 41 individual GCs
taken from Schiavon \etal(2005) whose Lick indices were measured and 
calibrated to the Lick system by Mendel, Proctor \& Forbes (2007), 
and 12 (Lick calibrated) GCs from Puzia \etal(2002b).
The Schiavon \etal(2005) and Puzia \etal(2002b) samples have 11 GCs
in common, giving a combined dataset of 53 spectra for 42 GCs. 
Since all the GC indices have been calibrated onto the 
Lick system, they are directly comparable to our data.

\begin{figure*}
\centerline{\includegraphics[width=12.0cm]{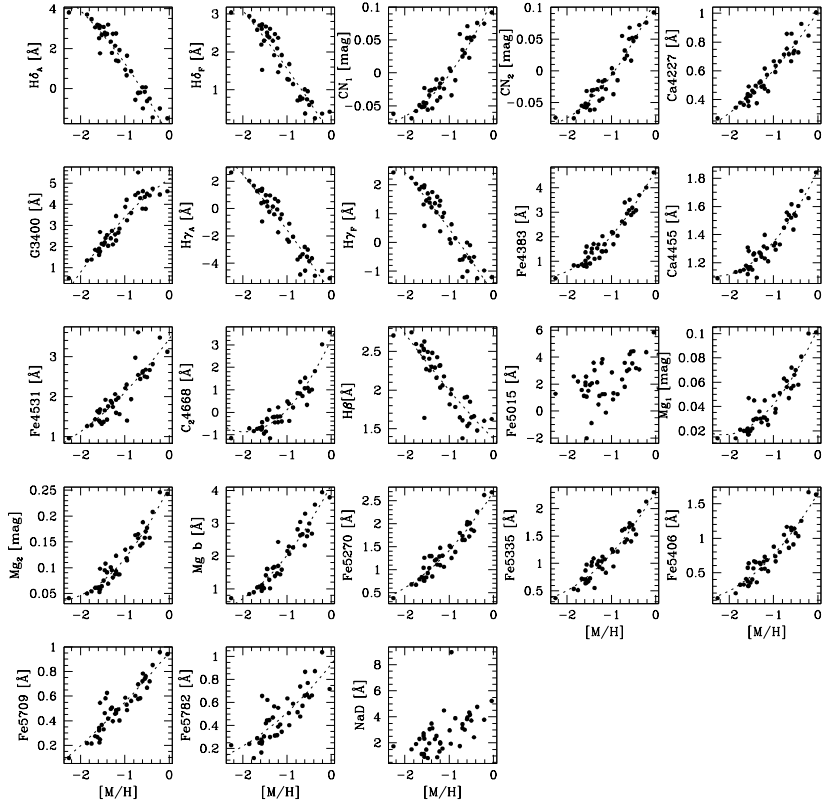} }
\caption{The behaviour of the Lick indices measured in the Schiavon \etal(2005)
sample as a function of the metallicities listed in the Harris catalogue.
Dashed lines are our least-squares polynomial fits to these data.
No fits where attempted for the Fe5015 and NaD indices.
\label{fig:MWrelations}}
\end{figure*}

 To derive empirical metallicities, we obtained correlations between the Lick 
index strengths and the Harris (1996, Feb 2003 version) catalogue metallicities 
of the Milky Way GCs (see e.g., Perret et al.~2002; Puzia et al.~2002b).
The majority of these relations appear non-linear and 
we therefore characterised the relations with second-order polynomial fits, 
giving a metallicity estimate for each index measurement. These relations are shown in
Figure~\ref{fig:MWrelations} and the polynomial coefficients and the
rms of the fits are listed in Table~\ref{tab:polynomials}.
All the indices, with the exceptions of Fe5015 and NaD, show
clear correlations with metallicity. The Fe5015 index is 
seemingly affected by poor sky subtraction (Schiavon \etal2005), whereas
NaD is possibly affected by interstellar absorption.
Of the remaining 21 available Lick indices we have chosen to use six
indices: Fe4383, Mg$_2$, Mg $b$, Fe5270, Fe5335 and Fe5406.
These indices all have high correlation coefficients (Table~\ref{tab:polynomials}), 
low residuals about the fit, and have relatively small uncertainties in 
the NGC 5128 data. 
We avoided the CN and G4300 indices due the unknown role of mixing 
in the GCs and also the Balmer lines which are affected 
by age variations. We also avoided the Mg$_1$ and Ca4455 indices which show 
signs of significant flattening at the lowest metallicities.

The individual index 
measurements were then combined employing a Tukey biweighting scheme 
(giving outliers less weight, although we note that weighted and unweighted 
averages yielded very similar results). Uncertainties were derived robustly using the 
median absolute deviation scaled to asymptotically 
approach a normal standard deviation (allowing the calculation of a standard error on the mean). 
Between our final metallicity estimates for the Milky Way GCs and the Harris (1996) values 
we found a scatter of 0.12 dex. We then proceeded to use the above approach to assign
metallicities to the NGC 5128 clusters based upon their Lick indices. Experiments suggested
that a S/N\footnote{Signal-to-noise (S/N) values in this study are defined per Angstrom,
and are the median value of the entire spectrum.} of 20 provided robust results, and we therefore 
took this as a (fairly conservative) minimum S/N for this exercise, leaving 207 GCs in the sample.

\begin{figure*}
\centerline{\includegraphics[width=14.0cm]{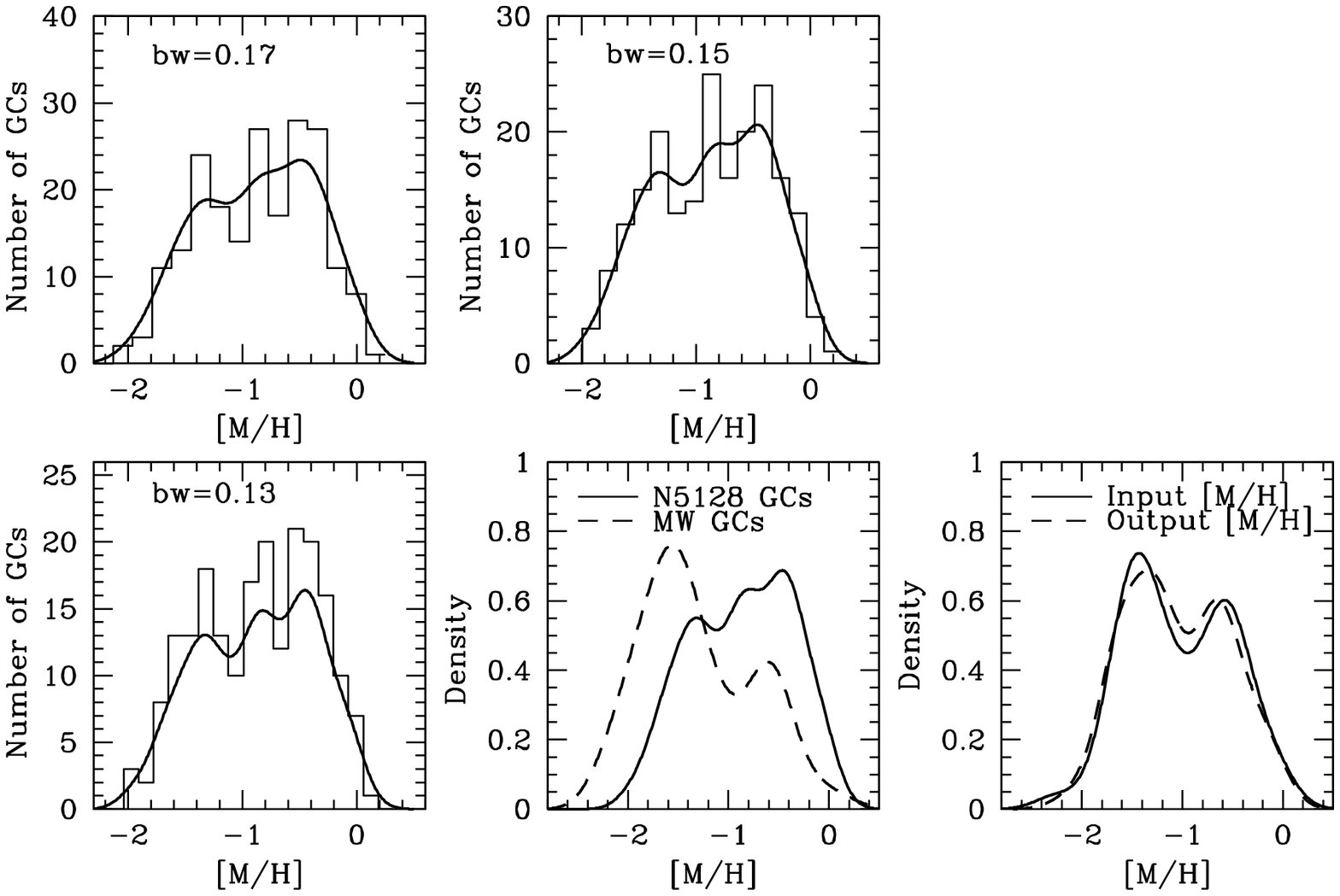} }
\caption{Metallicity distribution for 207 GCs (S/N$>20$) in NGC~5128 derived
from an empirical relation based on Milky Way GCs (see text).
{\it Top left panel:} histogram and Gaussian density kernel estimates 
with bin/kernel bandwidth  0.17 dex, 
{\it top centre panel:} 0.15 dex bin/kernel bandwidth, 
{\it bottom left panel:}  0.13 bin/kernel bandwidth (mean uncertainty in metallicities). 
{\it Bottom centre panel:} 
comparison between metallicity 
distributions of NGC~5128 GCs (solid line) and 148 Milky Way GCs (dashed line) taken 
from the Harris (1996) catalogue where the density kernel bandwidths are 0.15 and 0.19 for the 
NGC 5128 and the Milky Way GCs respectively.
{\it Bottom right panel:} Comparison of the input Milky Way GC metallicities
from the Harris catalogue with those recovered using the empirical metallicity calibration.
\label{fig:MH}}
\end{figure*}

This empirical metallicity scale is tied to the  Milky Way GC metallicities 
listed in the Harris (1996) catalogue
which comprise a mixture of Zinn \& West (1974) spectrophotometric metallicities 
and high resolution spectroscopic estimates. 
However, the high-dispersion spectroscopic measurements in
the literature are now numerous enough that for most Milky Way
clusters (including all the luminous, classic ones that we
use below, for example, for Lick index standardization)
these are the measurements that dominate the catalog metallicities.
What exactly this metallicity 
scale measures ([Fe/H], [Z/H] or something else) is a matter of debate 
(e.g., Thomas, Maraston \& Bender 2003; 
Mendel \etal 2007). Here we simply regard it 
as the ``Galactic GC metallicity scale''. To reflect this 
ambiguous nature, we refer to our empirical metallicities as simply [M/H], 
which are directly comparable to any other metallicities based upon those in the 
Harris catalogue. Furthermore, in the following sections we use the following notation: 
we refer to the Harris (1996) catalogue metallicities as [Fe/H], [Z/H] refers 
to our SSP-derived metallicities, [$\alpha$/Fe] signifies any literature estimates 
of the $\alpha$-element over iron ratio, and [E/Fe] is our SSP-derived ratio of 
''enhanced''($\alpha$) elements over iron.

The empirical metallicities and uncertainties for the NGC 5128 GCs
are listed in Table~\ref{tab:GCs}, and their resulting spectroscopic ``metallicity 
distribution function'' (MDF) is shown in Fig.~\ref{fig:MH}. Assuming a total population 
of 1500 GCs in NGC~5128 (Harris \etal2006b), this represents $\sim14\%$ of 
the GC system. To show the influence of our choice of binning on 
the precise form of the distribution
we show the MDF using three different bin widths of 0.17 dex, 0.15 dex 
and 0.13 dex. We also overplot nonparametric kernel density estimates 
using the same kernel bandwidths as the histogram bin widths\footnote{
For these data 0.17, 0.14 and 0.13 dex correspond approximately to Silverman's (1986)  
``rule of thumb'', the Sheather \& Jones (1991) method for bandwidth 
estimation and the median uncertainty in the empirical metallicities respectively.}.
Irrespective of the exact choice of binning or density kernel, the NGC~5128 GC MDF 
looks non-Gaussian. We tested for the presence of metallicity subpopulations
in the sample using KMM and the bayesian code Nmix (Richardson \& Green 1997; 
see discussion in Strader \etal2006). Using KMM in the 2-group, homoscedastic case, 
a bimodal distribution is preferred to a single gaussian with a $p$-value
of 0. Nmix also finds that a (heteroscedastic) bimodal distribution is more
probable than a unimodal one (45\% and $<1$\% probabilities respectively).
therefore, the GC system of NGC 5128 is convincingly multimodal
in {\it spectroscopic} metallicity, reflecting the existence of at least two
metallicity subpopulations (Strader \etal2007, Kundu \& zepf 2007) rather than
a single population (Yoon \etal2006).

It is also apparent that the empirical MDF of the NGC~5128 GCs is on average
more metal-rich than that of the Milky Way GCs.
A straight mean of the Milky Way GC metallicities gives [M/H]=--1.3 whereas 
the corresponding value for the NGC~5128 GCs is [M/H]=--0.87. 
This result seems to be driven by two factors. Firstly, there is a larger 
fraction of metal-rich (hereafter red) GCs in the 
NGC 5128 sample than in the Milky Way sample. 
Secondly, both the metal-poor (hereafter blue) GCs and red GCs in 
NGC 5128 are on average more metal-rich than the corresponding populations
in the Milky Way. In the bimodal case outlined above, KMM (Nmix)
finds peaks at [M/H]=--1.34 and --0.52 (--1.30, --0.50) for the NGC 5128 GCs, 
whereas the corresponding values for the Milky Way GCs 
are [M/H]=--1.62 and --0.61 (--1.62, --0.63). 
The blue GCs in NGC 5128 are $\sim0.3$ dex 
more metal-rich than the blue GCs in the Milky Way. The red GCs in NGC 5128 are also, 
on average, slightly more metal-rich than
the average of the Milky Way GCs ($\sim0.1$ dex).  

These differences in the average 
{\it spectroscopic} metallicities of the blue and red 
GC subpopulations between the two galaxies are broadly consistent 
with the positive correlation between galaxy luminosity and mean 
GC metallicity found photometrically (Larsen \etal2001; Strader \etal2004; Peng \etal2006). 
In detail, the blue peak of the NGC 5128 GCs lies close to the upper envelope
of the relation (Peng \etal2006; Brodie \& Strader 2006) -- i.e. more metal-rich than
the mean relation -- whereas the red peak lies close to the lower envelope of the 
red relation (i.e. more metal-poor than the mean relation).
Since the peaks in the MDF lie slightly off the galaxy 
colour-magnitude relations in such a fashion, no correction in the 
integrated magnitude of the galaxy can bring better agreement. 
However, both population peaks in our data lie within the 
scatter of the photometric relations.
 
We also note that the MDF the NGC 5128 GCs is {\it narrower} than that of the Milky Way GCs. 
The full-width half-maxima of the two distributions are 1.3 and 1.6 dex respectively.
This seems to be largely driven by the blue peak in the NGC 5128 MDF which, 
being more metal-rich on average than the Milky Way blue peak, has been pushed into the 
red peak of the MDF.

\begin{figure}
\centerline{\includegraphics[width=8.0cm]{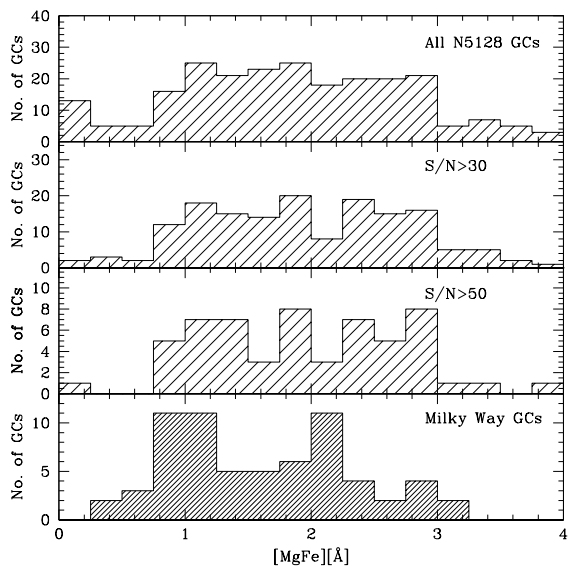} }
\caption{Distribution of [MgFe] of the NGC~5128 GCs for different S/N limits in 
the spectra. The lower panel shows the distribution for Milky Way GCs which is 
clearly bimodal. These NGC~5128 data exhibit a broad, perhaps trimodal distribution.
\label{fig:MgFe}}
\end{figure}

Inspection of the NGC 5128 MDFs in Figure~\ref{fig:MH} suggests that 
there may be further metallicity substructure in the GC system.
By eye, peaks in the MDF are apparent at [M/H]$\sim-1.3$, --0.4 and also 
perhaps at [M/H]$\sim-0.8$. Results for KMM in the trimodal case were 
inconclusive, but Nmix found peaks at --1.43, --0.87 and --0.38 with a 54\% likelihood
that there are three or more peaks in the distribution 
(recall likelihoods of 1\% and 45\% were found for the unimodal and bimodal cases
respectively.) A test to check on the reality of this ``trimodality'' is to 
see if this is evident in the raw Lick indices. In Figure~\ref{fig:MgFe} we show 
histograms of the [MgFe] index of the NGC~5128 GCs compared to the Milky Way sample 
used for the metallicity calibration. We show [MgFe] because this index has higher 
S/N than most individual indices (being a combination of three indices) and it is 
thought to be relatively insensitive to $\alpha$-abundance variations (Thomas \etal2003) 
The distribution of indices for all 254 GCs is broad and 
relatively flat. As the lower limit of the S/N of the spectra is increased, 
three peaks become increasingly evident. 
We note that similar behaviour is seen in other Lick indices with sufficient 
S/N, such as the Mg$_2$ feature. By contrast, the Milky Way sample 
(42 unique GCs) is clearly bimodal in [MgFe]. These results are suggestive that there may
be intrinsic substructure in the metal-rich peak of the N5128 MDF.

\begin{figure}
\centerline{\includegraphics[width=8.0cm]{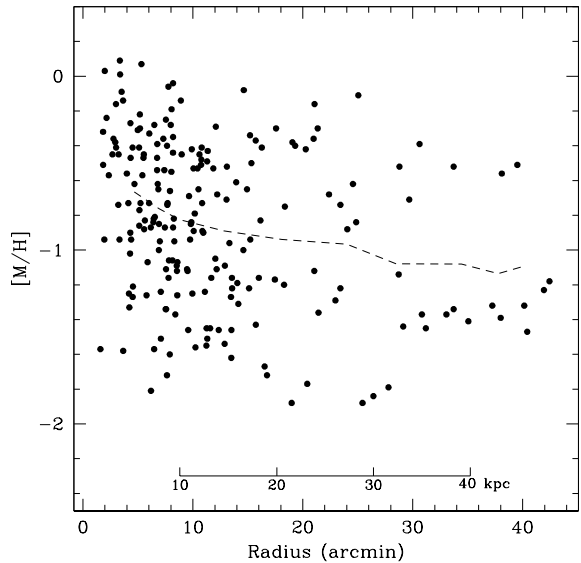} }
\caption{Radial distribution of spectroscopic metallicities for 207 GCs in NGC~5128.
The dashed line is a sliding mean using a width of 5 arcmin. 
\label{fig:radial}}
\end{figure}

The radial distribution of the GC metallicities is shown in Figure~\ref{fig:radial}.
The entire system shows a mild decrease in [M/H] with increasing radius as emphasized 
by the sliding-mean values. A linear fit to the GC data finds a gradient of 
--0.019 dex/arcmin ($\sim-0.017$ dex/kpc) with a large dispersion of $\sim0.5$ dex
(see also Woodley \etal2005). 
We find no significant gradients in either the blue or red GC subpopulations. 
This being the case, the metallicity gradient seen for the whole sample 
must be largely driven by an increasing fraction of blue/red GCs with radius 
(e.g., Perrett \etal2002; Harris \etal2006b).

One concern in using the Milky Way GCs as metallicity calibrators for extragalactic
GC systems is the small number of Milky Way GCs with high metallicities. In the combined
Schiavon+Puzia samples, only 5 unique GCs (10 spectra) have [Fe/H]$\geq-0.5$, with perhaps 
only NGC 6528 at solar metallicity.
In order to give the reader an idea of the uncertainties in our metallicity
calibration, we show in the bottom-right panel Fig.~\ref{fig:MH}
a comparison of the metallicities of the input Milky Way GC 
calibrators with those recovered using our calibration on these same data.
The plot shows that the extremes of the metallicity distribution are 
quite well constrained, although there is some small variation in the peak 
locations. KMM locates peak metallicities of --1.42 and --0.57 in the 
input sample, and finds peaks of --1.44 and --0.54 in the recovered
metallicities. The mean metallicity of the sample does not change.
Therefore, within the metallicity range of the Milky Way
GC calibrators, the empirical metallicity scale is fairly well constrained.
For metallicities above solar this is no longer the case. However, as shown
in Figure~\ref{fig:MgFe}, very few NGC 5128 GC have metallic indices stronger
than the most metal-rich Milky Way GCs, and therefore the exact behaviour
of the empirical calibration beyond solar metallicity is not of great concern.

\subsubsection{Comparison between GCs and field stars}
\label{ComparisonbetweenGCsandfieldstars}

The existence of star-by-star photometry of the giants in NGC 5128
allows us to compare directly the MDF of the field stars with that 
of its GCs. Photometric MDFs have been constructed 
at projected radii of 8 kpc (Harris \& Harris 2002), 
20 kpc (Harris \& Harris 2000) and at 38 kpc (Rejkuba \etal2005)
all determined from $(V-I)$ photometry 
of the halo red-giant branch (RGB) stars with the cameras on board {\sl HST}.
An obvious problem here is choosing at what radius the GCs should be 
compared with the field stars. In the absence of a compelling scientific criterium, 
we have chosen to compare our data with both the ``middle field'' at 
20 kpc ($\sim 4$R$_e$) from  Harris \& Harris (2000) and the ``inner field''
of Harris \& Harris (2002) at 8 kpc ($\sim 1.5$R$_e$).
Note that although both these field star MDFs are photometric and the
GC MDF is spectroscopic, all are calibrated to the metallicity scale of 
the Milky Way GCs in the Harris catalogue and are therefore 
readily comparable. 
We also note that our sample of GCs may be somewhat radially biased in the
sense that the metal-rich GCs may be overrepresented.
The KMM tests in performed in Section \ref{metallicityscale}
find a blue GC fraction of 41\% and a red GC fraction 
of 59\%. From a sample of 300 NGC 5128 GCs (of which 100 are from this
present study) Woodley \etal(2005) find that some 54\% are 
blue and 46\% are red. However, the true blue:red fraction will
have to await a complete photometric catalogue.

\begin{figure}
\centerline{\includegraphics[width=8.0cm]{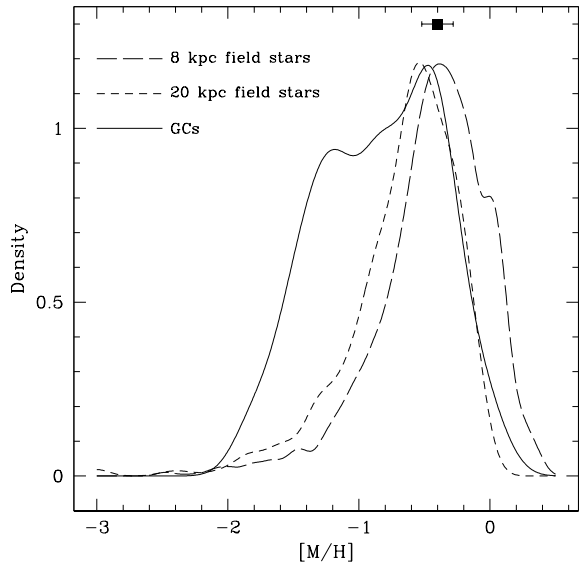} }
\caption{Comparison of the metallicity distribution functions of 
NGC 5128 halo field stars at 8 kpc (Harris \& Harris 2002) and 20 kpc 
(Harris \& Harris 2000) with the spectroscopic MDF of the NGC 5128 GCs. 
The normalised density 
distribution of the GCs has been scaled by $\times1.5$ to illustrate 
the similarities between the metal-rich ends of the distributions.
The solid square represents the metallicity derived from the r$_e/2$
integrated spectrum of NGC 5128 using our empirical metallicity calibration.
\label{fig:MDF}}
\end{figure}

The comparison of the MDFs is shown in Figure~\ref{fig:MDF}. 
In the figure, the normalised density distribution of the GCs 
has been scaled by a factor of 1.7 to illustrate the close 
coincidence of the metal-rich end of the GC MDF with that of the 20 kpc field
(at [M/H]$\sim-0.45$.)
In comparison, the average metallicity of the 8 kpc 
field star MDF is some 0.2 dex more metal rich than the GCs.
Some $\sim16\%$ of the RGB stars in the 8 kpc field are more metal-rich than
solar.
This is probably only a lower limit since incompleteness effects due to problems in 
dealing with the reddest RGB stars mean that the most metal-rich stars in the MDF are 
probably underrepresented (Harris \& Harris 2002). 
In addition, considering that the inner stellar field is at a galactocentric radius 
of $\sim 1.5$R$_e$, we might expect even more metal-rich stars in the central regions
of the galaxy than are seen in the 8 kpc MDF. 
In contrast, few of the GCs in our sample are convincingly
super-solar in metallicity; the most metal-rich GCs in NGC 5128 have comparable
metal-line strengths to the most metal-rich Milky Way GCs in the Schiavon \etal(2005) 
and Puzia \etal(2002) samples (NGC 6528 and NGC 6553). 
So, although our GC sample is by no means complete in the inner regions of NGC 5128, 
it presently appears that a significant fraction of the galaxy stars were able to obtain 
levels of heavy-element enrichment substantially higher than any of its GCs.

The other principal feature in the comparison of the MDFs is, 
as discussed by Harris \& Harris (2000; 2002), that the number ratio of GC
to stars varies considerably with metallicity.
No more than 10\% of the stellar MDF is more metal-poor than [M/H]$\sim-1$, 
whereas 40\% of the GCs are more metal-poor than [M/H]$\sim-1$. 
This implies that the ratio of efficiencies of metal-poor GCs to metal-poor stars 
is significantly higher than the corresponding metal-rich ratio, and/or 
the dynamical destruction efficiency of metal-poor GCs is considerably 
lower than that of their metal-rich counterparts.
From the r$_e/2$ integrated spectrum of the galaxy itself, we derive a
mean metallicity of [M/H]=$-0.40\pm0.12$, in agreement with the metallicities
found by Harris \& Harris (2002).

In constructing the NGC~5128 GC MDF with an empirical metallicity calibration 
tied to Milky Way GCs, we have made the implicit assumptions that the NGC~5128 
GCs have ages and abundance ratios comparable to the Milky Way GCs. 
These assumptions are tested in the following section where we 
estimate ages, metallicities and abundance ratios for a higher S/N subset of GCs
using SSP models.

\subsection{Stellar Population Models}
\label{SSPs}

Estimating ages, metallicities and abundance ratios from the integrated spectra 
of GCs requires the use of single-burst stellar 
population models (SSPs). However, as becomes readily apparent from
the extensive literature on the subject, such parameter estimates
remain model dependent. 

Motivated by the studies
of Proctor, Forbes \& Beasley (2004) and Mendel \etal(2007), who compared
SSP-derived ages, metallicities and abundance ratios with literature
estimates for Milky Way GCs, we used two different sets of SSPs.
We used the Lee \& Worthey (2005) models (henceforth LW05 models) and the 
Thomas \etal(2003) models, augmented with the higher order Balmer-line
calculations of Thomas, Maraston \& Korn (2004) (henceforth collectively 
referred to as TMK04 models). 
Mendel \etal(2007) showed that the LW05 models, 
when corrected for non-solar abundance ratios (NSARs) using the 
fractional sensitivities of Houdashelt \etal(2002; hereafter H02), and the 
TMK04 models are able to reproduce the ages, metallicities and [$\alpha$/Fe] ratios
for Milky Way GCs with [Fe/H]$\leq-0.5$ quite well. 
The LW05 models were particularly successful at recovering the 
[$\alpha$/Fe] patterns of Milky Way GCs found from high resolution 
spectroscopic estimates (e.g., Pritzl \etal2005).
At metallicities approaching the solar value, Mendel \etal(2007) found 
some evidence that the LW05 models underestimate the ages of MW GCs. 
In the case of the TMK04 SSPs, Mendel \etal(2007) showed that these models
predict an age--metallicity relation for Milky Way GCs in the opposite sense 
to those found by CMD determinations (De Angeli \etal2005). 
However, the TMK04 models seem better able 
to recover the ages of the most metal-rich ([Fe/H]$>-0.5$) Milky Way GCs.
We note that neither of these models has been well tested in the intermediate 
and young age regime. 

A detailed discussion of the LW05 models is given in LW05, and the correction
of these  models for NSARs is discussed in Mendel \etal(2007). 
Similarly, details of the TMK04 models are given in TMK04 and Thomas \etal(2003). 
For our purposes, the LW05 have been corrected for NSARs using the 
fractional sensitivities of H02, and have also 
been interpolated in age, metallicity and [E/Fe] to give finer grids spacings.
The TMK04 SSPs are used as published, but have also been interpolated 
in age, metallicity and [E/Fe]. TMK04 provide two sets of models which 
assume different amounts of mass loss on the horizontal branch. Here we use
the ``blue horizontal branch'' models, but note that this choice of model 
has no significant impact on our conclusions.

\subsubsection{Qualitative impressions}
\label{Qualitativeimpressions}

\begin{figure*}
\centerline{\includegraphics[width=16.0cm]{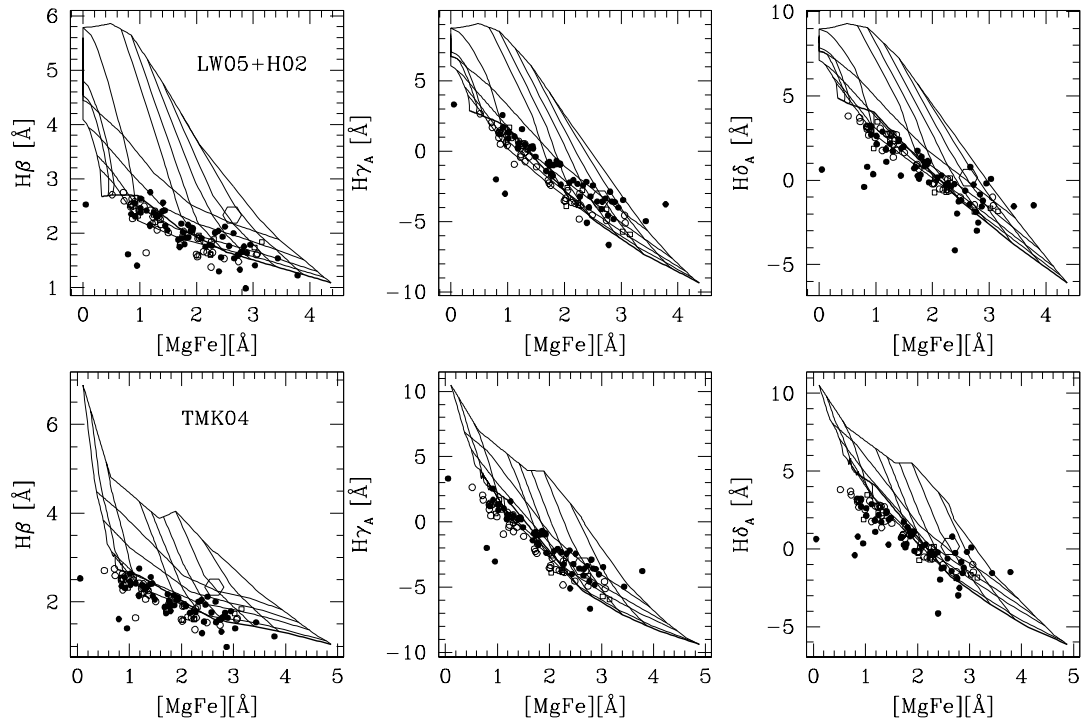} }
\caption{NGC~5128 and Milky Way GCs compared to SSP grids of LW05 (top panels) 
and TMK04 (lower panels) at [E/Fe]=0. 
Solid circles represent NGC~5128 GCs (S/N$\geq50$), open circles are MW GCs from 
Schiavon \etal(2005), squares are from Puzia \etal(2002).  
The large open hexagon shows the linestrengths 
of the NGC~5128 spheroid (r$_e/2$ aperture along the major axis).
The LW05 SSPs shown cover the parameter space ([Z/H]=--2.5,--2.2,--2,--1.5,--0.75,
--0.5,--0.25,0,0.25,0.5; ages=1,3,5,8,10,12 Gy) 
and TMK04 ([Z/H]=--2.25,--2.2,--2,--1.5,--0.75,--0.5,--0.25,0,0.2,0.5,0.67; 
ages=1,3,5,8,10,15 Gy). 
\label{fig:Grids1}}
\end{figure*}

In Figure~\ref{fig:Grids1} we show the NGC~5128 and Milky Way samples compared 
to the LW05 and TMK04 models at [E/Fe]=0. In the case of the NGC~5128 data, only 
clusters with median S/N$\geq50$ are shown in order to compare more directly with 
the higher S/N ($\gg50$) Milky Way data.
From the figure we see that the NGC~5128 and Milky Way GCs span a comparable range of 
metallicities and in general follow the old-age lines of the models. 
Notwithstanding the larger scatter in these NGC 5128 data, 
at [MgFe]$\sim2-3$ a number of the NGC~5128 GCs exhibit an enhancement in their Balmer
lines compared to the Milky Way sample. 
The locii of GCs with [MgFe]$<1.5$ coincide quite well in Figure~\ref{fig:Grids1}
suggesting that there are no gross offsets between the two GC samples (with the 
possible exception of H$\gamma_{A}$). This being the case, stronger Balmer indices in some
NGC 5128 GCs when compared to the Milky Way sample
(at a given [MgFe]) would conventionally be interpreted as 
some of the NGC~5128 clusters having younger ages than those in the Milky Way sample.
The position of the spheroid of NGC~5128 
indicates a luminosity-weighted near-solar metallicity and a younger age 
than the Milky Way GCs and the majority of the NGC 5128 GCs.

\begin{figure*}
\centerline{\includegraphics[width=16.0cm]{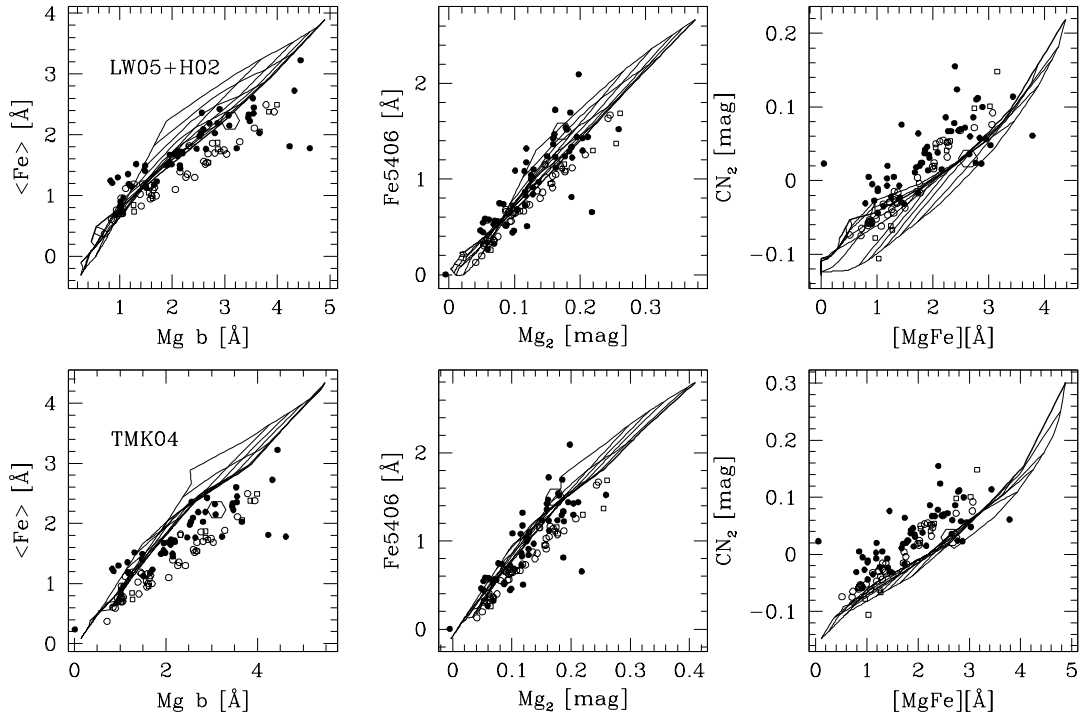} }
\caption{NGC~5128 and Milky Way GCs compared to SSP grids of LW05 (top panels) 
and TMK04 (lower panels) at [E/Fe]=0. Symbols and SSP grids are the same as 
for Fig.~\ref{fig:Grids1}. The NGC~5128 GCs on average exhibit lower Mg/Fe 
ratios than the MW GCs at a given metallicity. 
\label{fig:Grids2}}
\end{figure*}

Figure~\ref{fig:Grids2} shows Mg/Fe diagnostic model grids compared 
to the two datasets. In both the Mg$b$--$\langle$Fe$\rangle$ and Mg$_2$--Fe5406 
planes, the locii of the NGC~5128 GCs are shifted to the left of the Milky Way 
GCs (shifted to lower Mg and/or higher Fe). These plots suggest 
that the NGC~5128 GCs generally have lower Mg/Fe ratios than 
the Milky Way GCs at a given metallicity, and that the NGC 5128 spheroid
has lower Mg/Fe ratios than its GCs. In the right-hand panels of 
Figure~\ref{fig:Grids2} we compare the CN$_2$ measurements of the GCs at a 
fixed [MgFe]. The two GC systems show very similar CN values for a given [MgFe], 
and both systems seem strongly enhanced in CN with respect to the model grids and 
the spheroid light of NGC~5128. This CN enhancement is becoming a seemingly 
typical characteristic of GC systems (e.g., Cenarro \etal2007). 
We note that these NGC~5128 data show a significant spread in CN$_2$ when compared 
to the Milky Way GCs. 

\subsubsection{Multivariate model fits}
\label{Multivariatemodelfits}

To estimate the SSP model parameters which best describe these data, we employed 
the robust multi-index $\chi^2$ fitting approach detailed in Proctor \etal(2004). 
This method has been used with success to estimate ages, metallicities and 
abundance ratios for both galaxies and extragalactic GCs, giving consistent results 
with single index-index determinations, but is more robust when confronted with 
problems such as varying S/N and background emission (e.g., Proctor \etal2004; 
Pierce \etal2006).
Measured Lick index data are compared to the same indices measured from a SSP grid 
where [Z/H], age and [E/Fe] are allowed to vary. For each grid point a $\chi^2$ 
statistic is calculated. Any indices which are highly deviant from the fit are 
removed and $\chi^2$ recalculated (in our case, we ran three iterations with 
a 5-3-5$\sigma$ clipping pattern). 
The entire grid is then searched until $\chi^2$ is minimized 
(see Proctor \etal2004 for a detailed description).
Uncertainties on the three parameters were determined 
by randomly perturbing the individual indices by their uncertainties and recalculating
the fits.

In the case of the Milky Way data, the following indices were excluded from the 
entire fitting process : CN$_1$, CN$_2$ and Ca4227 (due to very strong C 
and/or N enhancement in the spectra), 
Fe4531 and Fe5015 (problematic sky-subtraction in the Schiavon \etal2005 data) and 
NaD (problems with interstellar absorption).
For the NGC~5128 data, indices excluded were CN$_1$, CN$_2$ and Ca4227, NaD, 
TiO$_1$ and TiO$_2$. The two TiO indices did not fall into the wavelength range 
of these data. We also excluded H$\gamma_{A}$ since there is some evidence of a 
systematic offset in this index compared to the Milky Way sample 
(e.g., see Figure~\ref{fig:Grids1}). 

\begin{figure}
\centerline{\includegraphics[width=8cm]{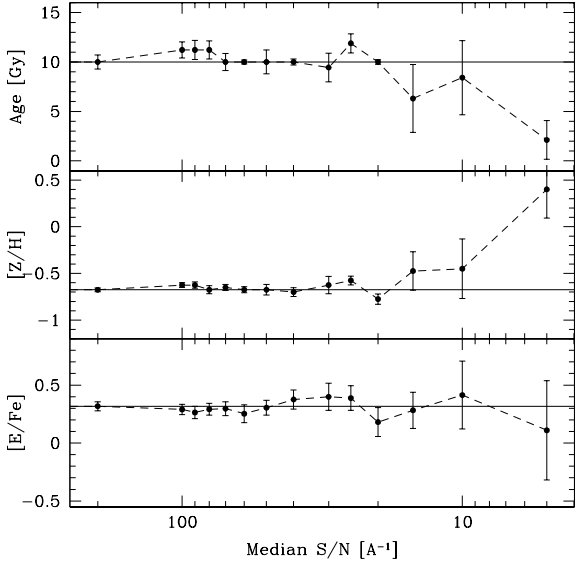} }
\caption{Ability of the $\chi^2$ fitting procedure to recover the input 
age, [Z/H] and [E/Fe] as a function of median S/N for the Milky Way GC
NGC~104 (47 Tuc). Symbols with error bars represent the median of 100 Monte Carlo
iterations with 1$\sigma$ uncertainties. The solid horizontal 
lines in each panel show the position of the initial best-fit 
parameter derived in this case using the LW05 SSP models.
\label{fig:Degrade}}
\end{figure}

Since the NGC~5128 spectra have a range of median S/N (10--114 \AA$^{-1}$) we tested 
the robustness of the fitting process as a function of the median S/N of the spectra. 
To this end we took the Schiavon \etal(2005) spectrum of NGC~104 (47 Tuc) 
and systematically degraded the S/N of the spectrum 
from 200--10 \AA$^{-1}$ by adding Poisson noise. For each S/N interval 100 
degraded spectra were created with a random seed used in each case. 
The $\chi^2$ fitting technique was then applied to the Lick indices measured 
from these spectra to see how well we could recover 
the initial best solutions. The results of this exercise are shown in 
Figure~\ref{fig:Degrade} using the LW05 models (similar results are obtained 
for the TMK04 models). Our simulations suggest that the $\chi^2$ approach 
recovers the initial parameters down to median S/N$\sim20-30$. 
At lower S/N we see that the fitting results become increasingly unreliable. 
In the example in Figure~\ref{fig:Degrade} it is noticeable that as the metallicity is 
overestimated, the age becomes increasingly underestimated which is a consequence of 
age--metallicity degeneracy. We also note that similar experiments on several 
other metal-rich GCs generally indicated a tendency to underestimate the age of 
the GCs at low S/N. Informed by our simulations we proceeded to choose a S/N of 
30 \AA$^{-1}$ as the lower limit for our SSP analysis. This cut in S/N
left us with a total of 147 GCs.

\begin{figure}
\centerline{\includegraphics[width=8.0cm]{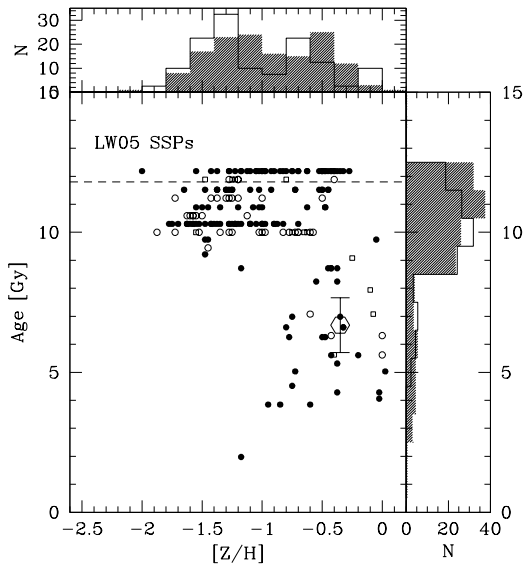} }
\caption{Ages and metallicities for all NGC~5128 GCs with S/N$>30$ (147 GCs)
and Milky Way data (symbols as for Figure~\ref{fig:Grids1}) using
the LW05 models.
The shaded and open histograms show the collapsed distributions in age and metallicity 
for the NGC~5128 and Milky Way GCs respectively. The NGC 5128 datapoints
have been offest by +0.3 Gy for display purposes.
The horizontal dashed line shows the old age limit of the models.
\label{fig:LWoresults} }
\end{figure}

\begin{figure}
\centerline{\includegraphics[width=8.0cm]{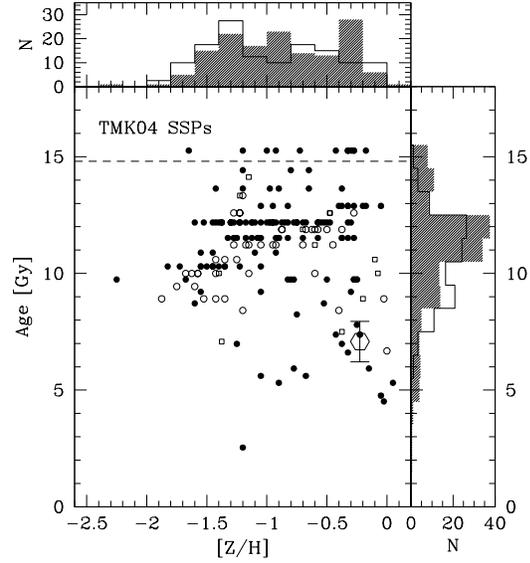} }
\caption{Ages and metallicities for NGC~5128 GCs 
and Milky Way data (symbols as for Figure~\ref{fig:Grids1}) using 
TMK04 models.
\label{fig:KMTresults} }
\end{figure}

The best LW05 and TMK04 model fits for age and metallicity for the two GC samples are 
shown in Figures~\ref{fig:LWoresults} and ~\ref{fig:KMTresults}. 
The corresponding (normalised) 
collapsed age and metallicity distributions are also shown. 
Error bars for the GCs have been omitted for clarity, 
but for example the mean errors in age and [Z/H] for the NGC 5128 GCs are
2 Gy and 0.14 dex respectively (at a mean S/N of 50).
For both sets of models, but more obviously for the 
LW05 SSPs, the NGC~5128 GCs appear to split into two groupings. 
There is an old ($>8$ Gy) population with a wide range
of metallicities and an apparently younger, predominantly metal-rich 
population ([Z/H]$>-1$). The results from using the LW05 models suggest that some 
22 out of 147 (15\%) of the GCs in our sample may be younger than 8 Gy
(henceforth, we will refer to these objects as intermediate
age cluster candidates, or IACCs). 
For the TMK04 models, this fraction drops to 17/147 (12\%)
of the sample. This disparity between the LW05 and TMK04 models in terms of 
the number of IACCs is a result of applying a fixed absolute age 
criteria (i.e. 8 Gy) when
there is a difference in the absolute age zeropoints between models 
(c.f. Figures~\ref{fig:LWoresults} and ~\ref{fig:KMTresults}).

Taken at face value, Figures~\ref{fig:LWoresults} and ~\ref{fig:KMTresults}
suggest that a significant
proportion of the {\sl metal-rich} GCs are distinctly younger than
the classic 12 to 13 Gy characterizing the metal-poor ones. 
In fact, the interpretation of the nature of these IACCs 
is complicated by the fact that a number of Milky Way GCs occupy a similar 
area of parameter space. Specifically these GCs are  NGC 6388, NGC 6441, 
NGC 6528, NGC 6553 and NGC 5927 (although note in that in 
Figures~\ref{fig:LWoresults} and ~\ref{fig:KMTresults}, these
clusters are over-represented since three of these clusters are in common 
between the Schiavon \etal(2005) and Puzia \etal(2002) samples).
NGC 6388 and NGC 6441 have been shown to have extended blue horizontal
branches despite their high metallicities (Rich \etal1997) which may be affecting 
their integrated light and thus biasing their age estimates. 
The other three clusters show no such obvious 
hot populations but still show enhanced Balmer lines. Understanding 
the origin of this young signature in the integrated light
of ostensibly old GCs is beyond the scope of this paper, but is clearly a priority.
Notwithstanding the presence of these ``younger'' Milky Way clusters, the age 
distributions of the two GC samples do appear statistically different. 
Kolmogorov-Smirnov (KS) tests comparing the 
age distributions of the NGC~5128 and Milky Way GCs find that the age distributions 
differ at 96\% (LW05 models) and 95\% confidence (TMK04 models).

We also identify one young cluster which appears to be an outlier in 
both figures. Cluster HGHH-G279 is younger and more metal-poor than the rest 
of the IACCs with [Z/H],age= -1.2, 1.7 Gy (-1.2, 2.2 Gy) LW05 models
(TMK04 models). The spectrum of this cluster is shown in Figure~\ref{fig:spectra}, 
which clearly shows the signature of a young stellar population.
HGHH-G279 was identified as a young cluster candidate spectroscopically
by Held \etal(2002) and photometrically by PFF04.

In the case of the integrated light of the galaxy itself, the r$_e/2$ bulge 
light occupies a position centred somewhere near the middle of the IACC 
subpopulation ($\sim6-7$ Gy, [Z/H]$\sim-0.3$). Our spectroscopic metallicity 
for the NGC 5128 bulge compares favourably with the determination of 
Harris \& Harris (2002) from direct {\it HST} photometry of red giants. These authors
determined an average metallicity [Fe/H]=--0.2 at a projected radius of 
8 kpc ($\sim1.6$ r$_e$) in NGC 5128. Our age estimates for the bulge are also in good 
agreement with literature determinations. Based on the AGB bump
and red clump locations from {\it HST}/ACS imaging of a 
more distant field (38 kpc projected radius), Rejkuba \etal2005 estimated an average
age of $8\pm3$ Gy for the halo stars.

\begin{figure}
\centerline{\includegraphics[width=8.0cm]{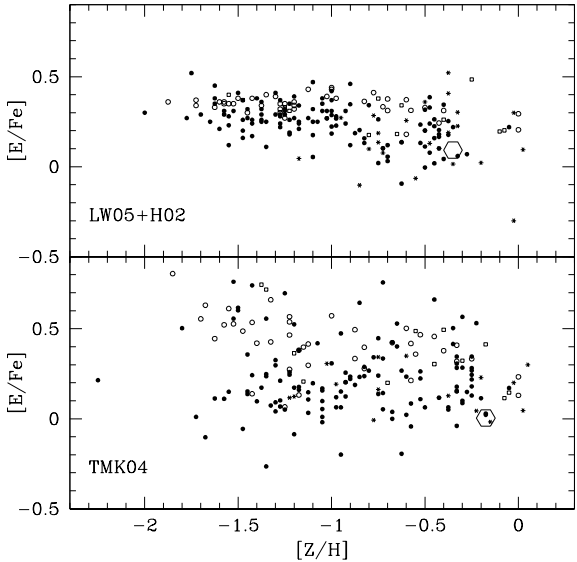} }
\caption{The run of [E/Fe] with metallicity for old NGC~5128 GCs (solid symbols), 
IACCs (asterisks) and the Milky Way sample (open symbols) from the SSP analysis.
\label{fig:Efe}}
\end{figure}

The behaviour of our [E/Fe] estimates versus [Z/H] derived from the LW05 and TMK04 
models for the two GC samples are shown in Figure~\ref{fig:Efe}.
Confirming the qualitative impressions given in 
Figure~\ref{fig:Grids2}, the NGC~5128 GCs appear to be, on average, offset to lower 
[E/Fe] than the Milky Way GCs at a given [Z/H]. This is true for both models and 
at all metallicities. However, the [E/Fe] predictions of the two SSPs
show significant differences. For the LW05 models, the NGC 5128 and 
Milky Way sample show quite similar behaviour, although the Milky Way GCs 
show a smaller scatter in [E/Fe] at a fixed [Z/H].

The [E/Fe] values derived using the TMK04 models are harder to interpret.
As shown by Mendel \etal(2007), the Milky Way GCs show a larger scatter
in [E/Fe] at a fixed metallicity than is the case for the LW05 models.
In addition, for the Milky Way sample [E/Fe] increases noticeably 
at [Z/H]$\leq-1$. However, the same behaviour is not seen in the NGC 5128 sample.
The NGC 5128 sample shows significant scatter, but the mean [E/Fe] remains roughly
constant with metallicity. This results in a situation where the locii of the 
two samples appear to increasingly diverge with decreasing metallicity.
The origin of this divergence a low metallicities is unclear, but is possibly related 
to the relative insensitivity of the TMK04 model indices to [E/Fe] at low metallicities.
Relatively small divergences between the model and data at low 
metallicities may be amplified and appear as large deviations in [E/Fe] as the models
``pinch'' together (see Mendel \etal2007).

In both models, the NGC~5128 GCs generally possess higher [E/Fe] than the galaxy bulge light ([E/Fe]$\sim0.1$). 
Separating by GC age, we find no statistically significant differences in [E/Fe] between 
the IACCs and old clusters.

\subsubsection{Comparison with PFF04}
\label{ComparisonwithPFF04}

PFF04 compared the Lick H$\beta$ index and [MgFe]' index
(defined by Thomas \etal2003) to the Thomas \etal2003 SSPs and determined an average 
age for the metal-rich GCs in NGC 5128 of 5$^{+3}_{-2}$ Gy. 
Of the 23 GCs in the PFF04 study, 15 are in common with our sample and with sufficient S/N
for age/metallicity analysis. Twelve of these we define as metal-rich ([Z/H]$>-1$).
We find a  mean age of these clusters of 8.9$\pm$2.7 Gy (LW05 models), 
roughly consistent with the PFF04 results. Of the IACCs in common between 
the samples, the three youngest GCs in PFF04 (HGHH-41, R261 and PFF-100) are also the youngest
GCs according to our analysis. Therefore, we find broad agreement between the two
studies. 

However, we do note an important distinction: PFF04 found that 7 of out 11 of their
{\it metal-rich} sample was younger than 8 Gy. If metal-rich GCs truly reflect field 
star formation, then this would imply that some 63\% of the galaxy field stars were 
formed at $z\lsim1$. Our results from the larger sample suggest an IACC fraction 
of somewhere between 20\% (TMK04 models) and 28\% (LW05 models).
However, this can at least in part be explained by the differences in the two samples.  
The GCs used in the PFF04 analysis were limited to the brightest GCs, all of which had
V$_0\leq 18.4$.  When we make the same magnitude cut on our larger sample, the
IACC fraction rises to 45\% of all metal-rich GCs.

\subsection{Other Properties of the Cluster System}
\label{Properties}

In the following we concentrate on results derived from the LW05 models which 
in general give lower reduced-$\chi^2$ values than the TMK04 models
and do not exhibit an age/metallicity trend for the old {\sl metal-poor} clusters.
However, this choice of model does not alter our basic conclusions. 

\begin{figure}
\centerline{\includegraphics[width=8.0cm]{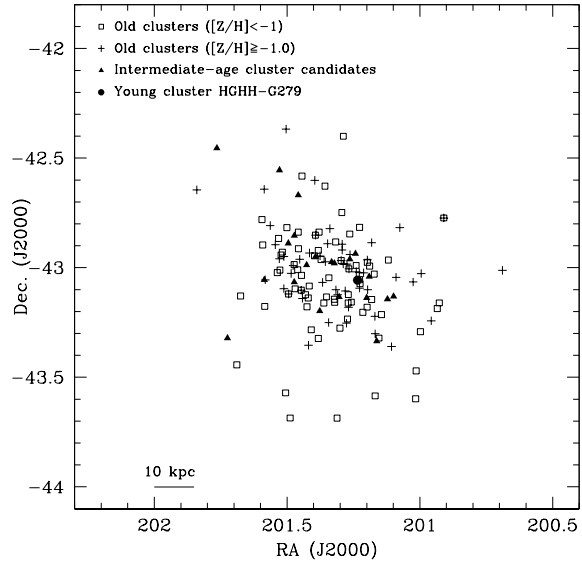} }
\caption{Spatial distribution of NGC~5128 GCs. 
\label{fig:Spatial}}
\end{figure}

The spatial distribution of the 147 NGC~5128 GCs which have both age and metallicity 
information is shown in Figure~\ref{fig:Spatial}. 
The IACCs show a clear preference towards the inner regions of the galaxy; 
their spatial clustering properties appear very similar to those of the old metal-rich clusters. 
In fact we see little difference between the spatial distributions of the 
IACCs, and the rest of the metal-rich sample.
The old blue clusters show a slightly more extended distribution, although 
the bias along the major axis in our spectroscopic sample and our incompleteness in the 
very central regions prevents a more quantitative analysis of these distributions
(the apparent ``ring'' feature is a result of this incompleteness in the centre).

\begin{figure}
\centerline{\includegraphics[width=5.0cm]{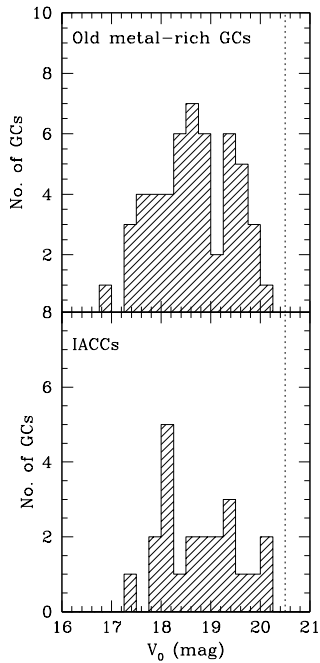} }
\caption{$V$-band luminosity functions of the old red GCs and the IACCs.
The vertical dashed line represents the position of the turnover in the luminosity function 
at V$_0\sim20.5$ (Harris \etal2004). 
\label{fig:LFs}}
\end{figure}

In Figure~\ref{fig:LFs}, we show the de-reddened $V$-band luminosity functions of the 
old red GCs and IACCs. The IACCs appear on average {\it fainter}
than the old red GCs, with means of V$_0=18.8\pm0.2$ and $18.2\pm0.5$ respectively.
However, this difference is not statistically significant; a KS test returns a 64\%
probability that the two distributions are drawn from the same parent distribution.

\section{Summary and Discussion}
\label{Summary}

We have performed a spectroscopic survey of GCs and GC candidates in the nearby elliptical
NGC 5128 using the 2dF instrument on the AAT. We obtained integrated optical 
spectra ($\lambda\lambda3800-6100$\AA) of 254 GCs of which 79 are newly confirmed
on the basis of their radial velocities and spectra. 
On higher S/N subsets of these data we derived empirical metallicities for 207 GCs and 
performed stellar population analyses using two different SSP models for 147 GCs. 
A parallel analysis was applied to a sample of 42 Milky Way GCs from Schiavon \etal(2005) and 
Puzia \etal(2002b), and 
to an r$_e/2$ aperture integrated spectrum of the bulge light constructed 
from 10 dedicated galaxy fibres. Our principal findings are:

\begin{itemize}

\item  The empirical spectroscopic metallicity distribution function (MDF) of the NGC 5128 
GCs is multimodal at high statistical significance, with two clear peaks at 
[M/H]$\sim-1.3$ and $\sim-0.5$.  
The mean metallicity of the ngc 5128 GCs is some 0.5 dex more metal-rich than 
that of the Milky Way GC sample.

\item There is evidence for a mild radial metallicity gradient in the NGC 5128 GC system 
($\sim-0.017$ dex/kpc). There is no evidence for gradients in either the metal-poor (blue) 
or metal-rich (red) subpopulations themselves; the global gradient is attributable to an
increasing ratio of blue to red GCs with radius.

\item A comparison between the MDFs of the NGC 5128 GCs and the field stars at 
20 kpc (Harris \& Harris 2000) reveals close coincidence in the locations of the metal-rich
ends of the distributions. However, the 8 kpc stellar MDF (Harris \& Harris 2002)
possesses a significant fraction of stars ($\geq16\%$) more metal-rich than the 
most metal-rich GCs in our sample. The ratio of GCs to field stars sharply increases with 
decreasing metallicity.

\item Comparison with stellar population models suggests that the majority 
of the NGC 5128 GCs ($\sim85-90\%$) have old ages ($>8$ Gy).
The remaining fraction of GCs appear younger than this, and all of these
objects lie in the red peak of the MDF. However, these GCs lie in a parameter
space occupied by a number of Milky Way GCs which have old ages derived from 
colour-magnitude diagrams, but have younger ages according to the 
SSP models. The presence of old, hot stellar populations 
in these GCs mimicking the effects of younger stellar populations cannot
be ruled out. The r$_e/2$ bulge light occupies a location roughly 
centred amongst the intermediate age cluster candidates (IACCs, $\sim6-7$ Gy old, 
[Z/H]$\sim-0.3$). 

\item SSP model predictions for the abundance ratios ([E/Fe]) show significant differences.
However, the [E/Fe] of the NGC 5128 GCs appear, on average, lower than
those of the Milky Way GCs at a given metallicity.  The IACCs are not differentiated 
from the old GCs in this regard. The abundance ratio of the bulge light is 
lower than the bulk of the GCs ([E/Fe]$\sim0.1$).

\end{itemize}

We begin by highlighting the principal uncertainty in our 
analysis. An outstanding, unsolved problem is how to deal with 
unresolved hot stellar populations (horizontal branch stars, blue stragglers etc.) 
in the integrated spectra of GCs, which can mimic the effect of young stellar 
populations (e.g., Burstein \etal1984). This is possibly the 
origin of the ``young'' SSP ages for the most metal-rich Milky Way 
GCs in this study, and remains a key uncertainty in 
attempts to age-date GCs from integrated spectra 
(e.g., Beasley \etal2002a; Puzia \etal2005; Cenarro \etal2007).
Hot stars in old stellar populations can and are, of course, included in SSPs, but in general
follow recipes based on Milky Way GCs 
(e.g., Vazdekis 1999; Maraston 2000; Lee \etal2000).
The behaviour of non-canonical populations cannot be modelled {\it a priori} 
and therefore must be estimated by observation. 
Wider wavelength ranges beyond 
the optical are urgently required to address this problem. In particular, we look 
forward to the use of {\it GALEX} in constraining the near and far UV flux and therefore 
the hot star populations in unresolved stellar systems.
This uncertainty in how to deal with non-canonical populations, and the differences in the age 
predictions of the TMK04 and LW05 models at high metallicity (see Section \ref{SSPs} and 
Mendel \etal(2007)), must be born in mind in the following discussion of GC ages.

One of the main results of this study is that the spectroscopic MDF 
of the NGC 5128 GCs is clearly bimodal, if not multimodal. 
The colour distribution
of this GC system also appears bimodal (e.g. Harris \etal2004), as do the 
colour distributions of GC systems for most massive galaxies (e.g., Larsen \etal2001; 
Harris \etal2004; Strader \etal2004; Peng \etal2006). 
Yoon \etal(2006) have suggested that such bimodal 
colour distributions might simply be a result of a non-linear colour-metallicity
relation, i.e., an intrinsically unimodal metallicity distribution can appear 
bimodal in broadband colours. However, our results indicate otherwise.
The MDFs of extragalactic GCs in massive galaxies {\it are} bimodal, and 
this needs to be accounted for by any models of GC system (and massive galaxy) 
formation. Our results for NGC 5128 are in agreement with the results
from IR photometry of M87 GCs (Kundu \& Zepf 2007) and with the Cohen \etal(2003)
spectroscopy of M49 GCs as shown by Strader \etal(2007).

We are again faced with the problem that the GC systems of massive 
galaxies are bimodal in colour/metallicity. 
From a chemical evolution point of view, 
a unimodal distribution is easier to understand than a bimodal 
one (e.g., Harris \& Harris 2002; Beasley \etal2003; Bekki \etal2003; 
VanDalfsen \& Harris 2004; Pipino \etal2007). This would be particularly
true if the MDFs of the GCs and field stars were to resemble each other, 
since the same star formation recipe may reasonably be expected to be
applicable to both. However, as shown in Section~\ref{Analysis}, 
the GC and stellar MDFs are quite different.
The most metal-rich ends of the distributions do show a strong resemblance
(c.f. Figure~\ref{fig:MDF}), but the radial gradient of the field stars in 
NGC 5128 suggests that the innermost stars in this galaxy are more metal-rich 
than the vast majority of the GCs found in this survey (Harris \& Harris 2002).
Moreover, any resemblance between the MDFs quickly disappears
with decreasing metallicity. The inference is that either the ratio
of GC formation to field star formation efficiency increases with
decreasing metallicity, or that the dynamical destruction efficiency
of GCs increases with increasing metallicity 
(or some combination of the two). This applies equally whether the GCs and 
field stars are formed {\it in situ} (e.g., Forbes \etal1997) 
or come from elsewhere (e.g., C{\^o}t{\'e}, Marzke \& West 1998; Beasley \etal2002b).
Understanding these differences, and indeed {\it how} exactly one should
compare the field stars to the GCs (e.g., at what radius) remains an
important aspect in understanding the galaxy--GC connection.

Our stellar population analysis suggests that there 
may be a mixture of GC ages in the NGC 5128 system. We find that 
some 10--15\% (SSP model dependent) of the sample may have intermediate ages 
($\sim4-8$ Gy). 
At least one cluster is younger and more metal-poor than the intermediate-age 
candidates (HGHH-G279; $\sim1-2$ Gy, [Z/H]$\sim-1.2$), 
with Peng \etal(2002) finding an even younger cluster ($<$ 1 Gy) in a
blue tidal stream in the halo of NGC 5128.
A $\sim15\%$ fraction of intermediate-age clusters is substantially lower than the 
$\sim67\%$ estimate claimed by PFF04, bringing NGC 5128 more into line
with the small contribution of intermediate-age populations found
in other nearby ellipticals (Strader \etal2005).
The difference between the present study and that of PFF04 can be understood 
largely in terms of the small sample size of the PFF04 study,
which with its brighter magnitude cut harboured an overrepresentative number 
of young clusters.

Beasley \etal(2003) specifically modelled the star formation histories
of NGC 5128--like GC systems (i.e., systems of similar mass, morphology 
and environmental density) with the semianalytical galaxy formation model 
of Cole \etal(2000).
Of the eight model systems presented in Beasley \etal(2003), the relative fractions
of metal-rich GCs formed within the past 8 Gy ($z\lsim1$) range from $\sim10-50\%$, with 
a median value of 20\%. If we define the fraction of young/old clusters in NGC 5128 
as only those GCs in the metal-rich peak ([Z/H]$>-1$), which one might expect are the GCs formed 
in later star formation episodes, we obtain a young cluster fraction of $\sim20-30\%$, 
in good agreement with the model predictions. 
Our results are also broadly consistent with the
``GC formation history'' derived by Kaviraj \etal(2005) who, from the $U-B$ colours
of 210 NGC 5128 GCs, estimated that some $25-35\%$ of GC 
mass was created 2--4 Gy ago.

\begin{figure}
\centerline{\includegraphics[width=8.0cm]{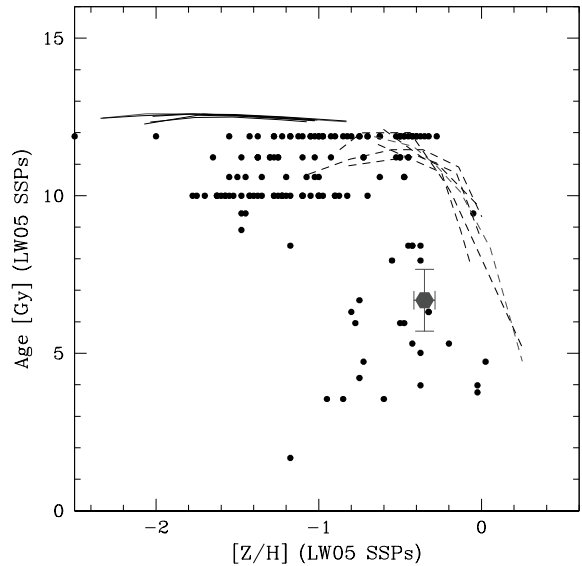} }
\caption{Ages and metallicities of 147 NGC 5128 GCs (derived from
LW05 SSP models) (solid circles) compared to the predicted age-metallicity relations
from Beasley \etal(2003). Solid lines are realisations for metal-poor GCs, 
the dashed lines are for the metal-rich, merger-formed GCs.
The hexagon with error bars represents the location of the 
r$_e/2$ bulge light.
\label{fig:AMR} }
\end{figure}

Beasley \etal(2003) also presented a series of age--metallicity relations
for GC systems formed using the semianalytic model. 
These relations are compared to the ages and metallicities 
of the GCs in our sample (LW05 models) in Figure~\ref{fig:AMR}.
At old ages, there is little discriminating power in the SSP models, and in the 
semianalytic the age of the metal-poor GCs is effectively fixed at $z=5$ 
the ``truncation redshift'' (corresponding to 12.3 Gy look-back time 
for the cosmology adopted in Beasley \etal2003).
For the metal-rich GCs, the model predicts a clear age-metallicity  
for the GCs (in fact, such a relation is a generic prediction of semi-analytic models).
The model age-metallicity relation reproduces the observations in the broadest sense 
(in the sense that there do appear to be young and metal-rich GCs in NGC 5128). 
However, two obvious problems remain. Firstly, it is not clear
whether the NGC 5128 sample shows a {\it relation} as such, but rather a large population of old GCs
and a smaller population of younger GCs with similar ages and metallicities. Indeed, the
spread in ages in the younger GCs is consistent with our observational errors
(although, the metallicities do show a large intrinsic spread).
Secondly, it is clear that the model age-metallicity relation is too metal-rich for the observations.
A reduction in the effective yield of the model (see Beasley \etal2003) would 
bring closer agreement with the observations, but this would result in the production 
of host galaxies which are too metal-poor for their luminosities, 
affecting the zeropoint of the colour-magnitude relation.

And here lies a generic problem 
with merger models which attempt to unify GC formation with the formation 
of the bulk of the host galaxy stars (e.g., Ashman \& Zepf 1992; Beasley et al. (2002); 
Bekki et al. 2002): in a model where metal-rich GCs and the galaxy stars are formed 
simultaneously through the same mechanism in mergers, the most metal-rich GCs should 
be as metal-rich as the most metal-rich galaxy stars. 
Comparison between the GC MDF and the 8 kpc stellar MDF (c.f. Figure~\ref{fig:MDF})
suggests that this does not seem to be the case since there is a clear excess of 
solar and super-solar metallicity stars with respect to GCs.
For this ``simple'' merger model to remain tenable, GC formation and star formation 
must be decoupled during the merger process such that the stars can enrich to significantly 
higher levels than the GCs. Otherwise, we are faced with the possibility that the 
mechanism which produces very metal-rich stars in galaxies is not that which produces 
metal-rich GCs.

\section{Acknowledgements}

MB thanks Trevor Mendel who provided the NSAR model corrections to 
the LW05 models and useful discussions, Javier Cenarro and Alex Vazdekis for discussions 
on the stellar population modelling, and Jay Strader for help on the empirical metallicity 
calibrations. WEH and GLHH thank the Natural Sciences and Engineering Research Council of Canada
for financial support.

\begin{table*}
\caption{Instrumental setup and observing details for 2dF spectroscopy.}
\label{tab:setup}
\begin{tabular}{@{}ll}
\hline
Telescope & 3.9-m AAT\\
Detector & Tek 1024$\times$1024 CCD\\
Image scale & 67\micron\ arcsec$^{-1}$\\
Pixel size & 24\micron\\
CCD inverse gain & 2.79 e$^-$ ADU$^{-1}$\\
CCD readout noise & 5.2 e$^-$\\
Dispersion & 2.4 \AA$^{-1}$ (spectrograph A), 1.8 \AA$^{-1}$ (spectrograph B)\\
Resolution${^1}$ & 5.3 \AA\ (spectrograph A), 4.0 \AA\ (spectrograph B)\\
Integration times & 25 hrs (3$\times$3 hrs shallow fields, 2$\times$8 hrs deep fields) \\
Seeing & 1.3--1.6 arcsec\\
\hline
\end{tabular}

\medskip
$^1$ some degradation of resolution with plate position - see text.
\end{table*}


\begin{table*}
\caption{Globular clusters in NGC 5128.}
\label{tab:GCs}
\begin{tabular}{@{}lccccccccc}
\hline
ID	& RA&	Dec.&	V & R$_{\rm V}$ & $\sigma_{\rm V}$ & Median S/N & RV template & [M/H] & $\sigma_{\rm [M/H]}$ \\
      &   (J2000)       &  (J2000)  &(mag)& (km/s)     & (km/s) & (\AA$^{-1}$) & Sp. type & (dex) & (dex) \\
\hline
AAT101906 & 13 23 58.76 & -43 01 35.2 & 18.9 & 511 & 31 & 55  &     K & 	-0.30 & 0.08 \\   								    
AAT101931 & 13 23 58.58 & -42 57 17.0 & 19.9 & 590 & 144 & 27  &    F &     	-1.93 & 0.17 \\  				
AAT102120 & 13 23 59.61 & -42 55 19.4 & 19.9 & 293 & 83 & 15  &     F &     	-2.15 & 0.48 \\  				
AAT103195 & 13 24 05.98 & -43 03 54.7 & 20.1 & 277 & 67 & 32  &     K &     	-0.47 & 0.18 \\  				
AAT106880 & 13 24 28.44 & -42 57 52.9 & 20.1 & 558 & 98 & 31  &     F &     	-1.66 & 0.24 \\  				
AAT107060 & 13 24 29.23 & -43 08 36.6 & 20.3 & 600 & 58 & 37  &     K &     	-0.59 & 0.13 \\
\end{tabular}
\end{table*}
\begin{table*}
\contcaption{}
\begin{tabular}{@{}lccccccccc}
\hline
ID	& RA&	Dec.&	V & R$_{\rm V}$ & $\sigma_{\rm V}$ & Median S/N & RV Template & [M/H] & $\sigma_{\rm [M/H]}$ \\
      &   (J2000)       &  (J2000)  &(mag)& (km/s)     & (km/s) & (\AA$^{-1}$) & Sp. type & (dex) & (dex) \\
\hline				
AAT204119 & 13 25 25.70 & -42 37 40.9 & 19.8 & 404 & 74 & 33  &     G &     	-1.36 & 0.21 \\  				
AAT205071 & 13 25 30.74 & -42 30 16.0 & 18.9 & 288 & 61 & 20  &     G &     	-1.57 & 0.52 \\  				
AAT208206 & 13 25 46.49 & -42 34 54.1 & 19.5 & 293 & 115 & 51  &    F &     	-1.87 & 0.19 \\  				
AAT209412 & 13 25 53.30 & -42 30 52.7 & 20.2 & 998 & 250 & 10  &    G &     	-1.28 & 1.00 \\  				
AAT215171 & 13 26 20.66 & -42 38 32.0 & 19.7 & 527 & 48 & 33  &     K &     	-0.57 & 0.27 \\  				
AAT222977 & 13 26 58.91 & -42 38 53.4 & 20.3 & 655 & 112 & 25  &    F &     	-1.30 & 0.45 \\
\end{tabular}
\end{table*}
\begin{table*}
\contcaption{}
\begin{tabular}{@{}lccccccccc}
\hline
ID	& RA&	Dec.&	V & R$_{\rm V}$ & $\sigma_{\rm V}$ & Median S/N & RV Template & [M/H] & $\sigma_{\rm [M/H]}$ \\
      &   (J2000)       &  (J2000)  &(mag)& (km/s)     & (km/s) & (\AA$^{-1}$) & Sp. type & (dex) & (dex) \\
\hline				
HGHH-G342 & 13 25 05.83 & -42 59 00.6 & 18.2 & 566 & 30 & 41  &     K &     	-0.43 & 0.15 \\  				
HGHH-G348 & 13 25 11.10 & -42 58 03.0 & 18.9 & 330 & 61 & 44  &     G &     	-0.94 & 0.10 \\  				
HGHH-G359 & 13 25 32.42 & -42 58 50.2 & 18.9 & 551 & 53 & 14  &     K &     	-2.41 & 1.76 \\  				
HGHH-G369 & 13 24 57.52 & -42 59 23.3 & 18.7 & 558 & 41 & 59  &     G &     	-1.33 & 0.07 \\  				
HGHH-G370 & 13 25 42.25 & -42 59 17.0 & 18.4 & 557 & 57 & 56  &     G &     	-0.81 & 0.13 \\  				
HGHH-G378 & 13 25 46.13 & -43 01 22.9 & 19.5 & 515 & 43 & 29  &     K &     	-0.01 & 0.13 \\
\end{tabular}
\end{table*}
\begin{table*}
\contcaption{}
\begin{tabular}{@{}lccccccccc}
\hline
ID	& RA&	Dec.&	V & R$_{\rm V}$ & $\sigma_{\rm V}$ & Median S/N & RV Template & [M/H] & $\sigma_{\rm [M/H]}$ \\
      &   (J2000)       &  (J2000)  &(mag)& (km/s)     & (km/s) & (\AA$^{-1}$) & Sp. type & (dex) & (dex) \\
\hline
PFF-gc014 & 13 24 41.05 & -42 59 48.4 & 20.2 & 718 & 59 & 19  &     K &     	-1.27 & 0.22 \\  				
PFF-gc015 & 13 24 41.20 & -43 01 45.6 & 19.2 & 571 & 44 & 45  &     G &     	-1.06 & 0.12 \\  				
PFF-gc016 & 13 24 43.60 & -42 53 07.3 & 19.9 & 521 & 46 & 32  &     K &     	-0.32 & 0.10 \\  				
PFF-gc018 & 13 24 47.10 & -43 06 01.7 & 18.9 & 531 & 29 & 66  &     K &     	-0.09 & 0.05 \\  				
PFF-gc022 & 13 24 54.33 & -43 03 15.5 & 19.8 & 601 & 75 & 32  &     G &     	-1.42 & 0.45 \\  				
PFF-gc023 & 13 24 54.55 & -42 48 58.7 & 19.4 & 504 & 53 & 46  &     G &     	-1.19 & 0.25 \\
\end{tabular}
\end{table*}
\begin{table*}
\contcaption{}
\begin{tabular}{@{}lccccccccc}
\hline
ID	& RA&	Dec.&	V & R$_{\rm V}$ & $\sigma_{\rm V}$ & Median S/N & RV template & [M/H] & $\sigma_{\rm [M/H]}$ \\
      &   (J2000)       &  (J2000)  &(mag)& (km/s)     & (km/s) & (\AA$^{-1}$) & Sp. type & (dex) & (dex) \\
\hline
WHH-18 & 13 25 30.08 & -42 56 47.0 & 18.4 & 721 & 40 & 66  & 	    K &     	-0.47 & 0.12 \\  				
WHH-20 & 13 25 34.36 & -42 51 06.0 & 19.0 & 269 & 42 & 51  & 	    G &     	-0.93 & 0.04 \\  				
WHH-25 & 13 25 50.37 & -43 04 08.3 & 20.0 & 515 & 58 & 25  & 	    K &        	-0.73 & 0.23 \\  				
\hline
\end{tabular}

\medskip
GC nomenclature: AAT= this study; HCH=Holland, C{\^o}t{\'e} \& Hesser (1999);
HGHH=Harris \etal(1992); HHH86=Hesser, Harris \& Harris (1986); 
HH=Harris \& Harris (2004); K=Kraft \etal(2001); MAGJM=Minniti \etal(1996); 
R=Rejkuba (2001); PFF=Peng \etal(2004); WHH=Woodley \etal(2005)
\end{table*}

\begin{table*}
\caption{Objects classified as foreground stars.}
\label{tab:stars}
\begin{tabular}{@{}lccccclccccc}
\hline
ID	& RA&	Dec. & V & R$_{\rm V}$ & $\sigma_{\rm V}$ & ID	& RA&	Dec.& V & R$_{\rm V}$ & $\sigma_{\rm V}$ \\
      &  (J2000)        & (J2000)   &  (mag) & (km/s)& (km/s) &       &  (J2000)        &   (J2000)&  (mag) & (km/s) & (km/s) \\
\hline
AAT101585 & 13 23 57.23 & -42 53 28.0 & 17.9 & 244 & 58 &     AAT230032 & 13 27 32.58 & -42 34 12.3 & 19.3 & 35 & 57\\  
AAT102095 & 13 24 00.48 & -42 44 55.0 & 18.1 & 158 & 65 &     AAT230068 & 13 27 32.81 & -42 27 50.9 & 18.3 & 232 & 76\\ 
AAT102515 & 13 24 02.59 & -42 46 44.1 & 19.7 & 126 & 127 &    AAT230095 & 13 27 32.91 & -42 24 05.0 & 18.5 & -8 & 37\\  
AAT103919 & 13 24 10.23 & -43 16 54.0 & 20.3 & 57 & 78 &      AAT233409 & 13 27 49.47 & -42 37 44.1 & 18.4 & -53 & 40\\ 
AAT104121 & 13 24 12.74 & -42 44 51.3 & 19.5 & 140 & 138 &    AAT233795 & 13 27 49.53 & -42 16 11.4 & 20.0 & 83 & 81\\  
AAT105234 & 13 24 19.31 & -42 49 03.1 & 20.1 & 177 & 133 &    AAT241043 & 13 28 10.08 & -42 44 52.9 & 20.1 & -60 & 87\\
\end{tabular}
\end{table*}
\begin{table*}
\contcaption{}
\begin{tabular}{@{}lccccclccccc}
\hline
ID	& RA&	Dec. & V & R$_{\rm V}$ & $\sigma_{\rm V}$ & ID	& RA&	Dec.& V & R$_{\rm V}$ & $\sigma_{\rm V}$ \\
      &  (J2000)        & (J2000)    &  (mag) & (km/s)& (km/s) &       &  (J2000)        &   (J2000)   &  (mag) & (km/s) & (km/s) \\
\hline
AAT223804 & 13 27 03.92 & -42 31 38.0 & 20.0 & 34 & 102 &     K-122 & 13 25 32.16 & -43 10 41.0 & 19.2 & 23 & 57\\	  
AAT224771 & 13 27 08.24 & -42 16 20.0 & 19.8 & -30 & 68 &     K-192 & 13 25 48.71 & -43 03 23.4 & 18.7 & 27 & 45\\	  
AAT227692 & 13 27 21.49 & -42 25 59.4 & 20.3 & 15 & 208 &     K-200 & 13 25 53.63 & -43 01 33.0 & 19.0 & 197 & 201\\	  
AAT228436 & 13 27 25.03 & -42 36 53.2 & 20.3 & -73 & 159 &    K-209 & 13 25 57.42 & -42 53 41.6 & 19.1 & 133 & 176\\	  
AAT230032 & 13 27 32.58 & -42 34 12.3 & 19.3 & 35 & 57 &      PFF-gc054 & 13 25 30.28 & -43 41 53.6 & 20.0 & 169 & 108\\
\hline
\end{tabular}
\medskip
\end{table*}

\begin{table*}
\caption{Objects classified as background galaxies.}
\label{tab:galaxies}
\begin{tabular}{@{}lccclccc}
\hline
ID	& RA &	Dec. &	 R$_{\rm V}$ & ID	& RA &	Dec.&	 R$_{\rm V}$ \\ 
      &  (J2000) &  (J2000)  &  ($10^{4}$ km/s) & & (J2000) &  (J2000)  &  ($10^{4}$ km/s) \\
\hline
AAT102536 & 13 24 02.23 & -42 59 48.3 & 5.9 & AAT335231 & 13 26 25.17 & -43 25 35.1 & 5.3\\
AAT103264 & 13 24 06.92 & -43 03 45.8 & 6.3 & AAT337834 & 13 26 36.38 & -43 22 21.2 & 3.5\\
AAT103993 & 13 24 11.42 & -43 14 25.2 & 3.4 & AAT344178 & 13 26 56.27 & -43 31 30.4 & 4.9\\
AAT104679 & 13 24 16.00 & -43 04 32.7 & 3.8 & HGHH-27 & 13 26 37.99 & -42 45 49.9 & 0.2\\  
AAT105959 & 13 24 23.99 & -42 54 10.8 & 4.6 & HH-001 & 13 24 02.67 & -42 48 32.2 & 0.5\\   
AAT106695 & 13 24 28.18 & -42 53 04.3 & 1.3 & HH-004 & 13 24 30.88 & -42 52 40.9 & 3.5\\
\hline
\end{tabular}
\medskip
\end{table*}

\clearpage

\begin{table*}
\caption{Lick indices measured for NGC~5128 globular clusters.}
\label{tab:LickIndices}
\begin{tabular}{@{}lccccccccccc}
\hline
ID & H$\delta_{\rm A}$ & H$\delta_{\rm F}$ & CN$_1$ & CN$_2$ &  Ca4227 & G4300 & H$\gamma_{\rm A}$ & H$\gamma_{\rm F}$ & Fe4383 & Ca4455 & Fe4531\\ 
   & (\AA) & (\AA) &  (mag) &  (mag) &  (\AA) &  (\AA) &  (\AA) &  (\AA) &  (\AA) &  (\AA) &  (\AA) \\
\hline
AAT101906 & 0.785 & 1.218 & 0.042 & 0.060 & 0.743 & 4.451 & -3.595 & -0.671 & 3.707 & 1.164 & 3.499\\
AAT103195 & -2.774 & -0.416 & -0.100 & -0.073 & 1.537 & 5.786 & -5.228 & -3.009 & 1.893 & 0.595 & 1.030\\
AAT106880 & 0.805 & 0.355 & -0.052 & -0.071 & 0.663 & 2.125 & 0.915 & 1.622 & 0.326 & 1.625 & 0.185\\
AAT107060 & 1.139 & 1.038 & -0.022 & 0.023 & 1.163 & 4.021 & -2.474 & 0.164 & 2.078 & 0.153 & 1.218\\
AAT107977 & 1.725 & 1.445 & -0.072 & -0.035 & 0.681 & 2.304 & 0.116 & 0.394 & -0.307 & 0.231 & 1.624\\
AAT109380 & -0.404 & 0.520 & -0.045 & -0.031 & 0.231 & 3.770 & -2.003 & -0.174 & 1.777 & 0.469 & 1.221\\
\end{tabular}
\end{table*}
\begin{table*}
\contcaption{}
\begin{tabular}{@{}lccccccccccc}
\hline
ID & H$\delta_{\rm A}$ & H$\delta_{\rm F}$ & CN$_1$ & CN$_2$ &  Ca4227 & G4300 & H$\gamma_{\rm A}$ & H$\gamma_{\rm F}$ & Fe4383 & Ca4455 & Fe4531\\ 
   & (\AA) & (\AA) &  (mag) &  (mag) &  (\AA) &  (\AA) &  (\AA) &  (\AA) &  (\AA) &  (\AA) &  (\AA) \\
\hline
HGHH-G219 & -1.482 & 0.351 & 0.043 & 0.061 & 0.799 & 4.262 & -3.775 & -0.479 & 4.009 & 0.993 & 3.880\\
HGHH-G251 & 0.232 & 1.300 & 0.016 & 0.044 & 0.351 & 4.420 & -5.186 & -1.253 & 3.939 & 0.768 & 3.363\\
HGHH-G277 & 1.782 & 2.040 & -0.017 & 0.008 & 0.535 & 2.661 & 0.126 & 0.617 & 1.914 & 0.857 & 1.242\\
HGHH-G279 & 5.133 & 3.901 & -0.142 & -0.173 & -0.274 & -2.198 & 6.235 & 4.374 & -0.477 & 0.368 & 0.852\\
HGHH-G293 & 3.825 & 2.239 & -0.061 & -0.035 & 0.625 & 2.675 & 0.391 & 0.839 & 1.150 & 0.620 & 1.041\\
HGHH-G342 & -1.172 & 1.116 & 0.071 & 0.123 & 0.980 & 4.837 & -3.556 & -0.316 & 5.000 & 1.177 & 3.977\\
\end{tabular}
\end{table*}
\begin{table*}
\contcaption{}
\begin{tabular}{@{}lccccccccccc}
\hline
ID & H$\delta_{\rm A}$ & H$\delta_{\rm F}$ & CN$_1$ & CN$_2$ &  Ca4227 & G4300 & H$\gamma_{\rm A}$ & H$\gamma_{\rm F}$ & Fe4383 & Ca4455 & Fe4531\\ 
   & (\AA) & (\AA) &  (mag) &  (mag) &  (\AA) &  (\AA) &  (\AA) &  (\AA) &  (\AA) &  (\AA) &  (\AA) \\
\hline
PFF-gc069 & 0.477 & 2.232 & 0.096 & 0.124 & 0.976 & 6.490 & -3.030 & -0.070 & 2.919 & 1.236 & 2.047\\
PFF-gc074 & 2.453 & 2.922 & -0.036 & -0.011 & 0.666 & 3.721 & -0.638 & 1.465 & 2.777 & 0.798 & 2.139\\
PFF-gc075 & -0.230 & 0.968 & 0.000 & 0.028 & 0.844 & 4.833 & -3.362 & -0.273 & 4.638 & 0.597 & 4.044\\
PFF-gc079 & 1.696 & 2.474 & -0.045 & -0.038 & 0.409 & 3.931 & -2.148 & -0.290 & 2.459 & 0.851 & 2.495\\
PFF-gc081 & 2.289 & 2.222 & -0.009 & 0.015 & 0.519 & 2.418 & 0.724 & 1.199 & -0.246 & 0.762 & 1.004\\
PFF-gc082 & -3.192 & -1.235 & 0.097 & 0.093 & 0.857 & 3.323 & -2.707 & -0.592 & 2.536 & 1.306 & 3.842\\
\end{tabular}
\end{table*}
\begin{table*}
\contcaption{}
\begin{tabular}{@{}lcccccccccccc}
\hline
ID & C$_2$4668 & H$\beta$ & Fe5015 & Mg$_1$ & Mg$_2$ & Mg $b$ & Fe5270 & Fe5335 & Fe5406 & Fe5709 & Fe5782 & Na D\\
   & (\AA) & (\AA) & (\AA) &  (mag) &  (mag) &  (\AA) &  (\AA) &  (\AA) &  (\AA) &  (\AA) &  (\AA) & (\AA) \\
\hline
AAT101906 & 2.396 & 1.538 & 4.222 & 0.061 & 0.185 & 3.070 & 2.484 & 2.150 & 1.695 & 0.667 & 0.572 & 1.124\\
AAT103195 & -1.130 & 0.258 & 2.546 & 0.085 & 0.263 & 4.235 & 2.227 & 1.355 & 1.385 & 0.319 & -0.206 & 0.539\\
AAT106880 & -0.617 & 2.715 & 1.815 & 0.041 & 0.080 & 0.958 & 0.434 & 0.298 & 1.011 & 0.314 & -0.068 & -0.694\\
AAT107060 & 1.921 & 1.881 & 4.190 & 0.069 & 0.168 & 3.445 & 2.222 & 1.318 & 1.395 & 0.911 & 0.105 & 6.539\\
AAT107977 & -0.119 & 2.040 & 1.913 & 0.053 & 0.127 & 1.968 & 1.615 & 1.813 & 1.308 & 0.441 & 0.447 & 2.768\\
AAT109380 & -0.980 & 1.614 & 2.339 & 0.017 & 0.052 & 0.929 & 0.825 & 0.531 & 0.545 & 0.270 & -0.039 & -0.081\\
\end{tabular}
\end{table*}
\begin{table*}
\contcaption{}
\begin{tabular}{@{}lcccccccccccc}
\hline
ID & C$_2$4668 & H$\beta$ & Fe5015 & Mg$_1$ & Mg$_2$ & Mg $b$ & Fe5270 & Fe5335 & Fe5406 & Fe5709 & Fe5782 & Na D\\
   & (\AA) & (\AA) & (\AA) &  (mag) &  (mag) &  (\AA) &  (\AA) &  (\AA) &  (\AA) &  (\AA) &  (\AA) & (\AA) \\
\hline
HGHH-G204 & -0.313 & 2.614 & 0.608 & 0.019 & -0.003 & -0.634 & -0.377 & 0.657 & 0.275 & 0.036 & -0.242 & 1.809\\
HGHH-G219 & 4.053 & 1.225 & 5.544 & 0.045 & 0.198 & 4.441 & 3.148 & 3.300 & 2.093 & 1.212 & 0.839 & 3.859\\
HGHH-G251 & 3.921 & 2.377 & 4.896 & 0.045 & 0.186 & 4.088 & 2.162 & 2.752 & 1.902 & 0.900 & 0.632 & 2.405\\
HGHH-G277 & 0.930 & 2.332 & 3.273 & 0.025 & 0.115 & 1.422 & 1.789 & 1.223 & 0.839 & 0.168 & 0.190 & -0.373\\
HGHH-G279 & 0.695 & 4.323 & 2.194 & 0.013 & 0.050 & 0.226 & 1.183 & 0.491 & 1.002 & 0.164 & 0.089 & 2.842\\
HGHH-G293 & 1.015 & 1.333 & 3.772 & 0.006 & 0.065 & 1.281 & 1.273 & 1.295 & 0.560 & 0.433 & 0.291 & -1.578\\
\end{tabular}
\end{table*}
\begin{table*}
\contcaption{}
\begin{tabular}{@{}lcccccccccccc}
\hline
ID & C$_2$4668 & H$\beta$ & Fe5015 & Mg$_1$ & Mg$_2$ & Mg $b$ & Fe5270 & Fe5335 & Fe5406 & Fe5709 & Fe5782 & Na D\\
   & (\AA) & (\AA) & (\AA) &  (mag) &  (mag) &  (\AA) &  (\AA) &  (\AA) &  (\AA) &  (\AA) &  (\AA) & (\AA) \\
\hline
PFF-gc069 & 3.492 & 1.254 & 5.470 & 0.043 & 0.190 & 3.116 & 2.796 & 2.698 & 0.962 & 0.805 & 0.438 & 3.435\\
PFF-gc074 & 1.795 & 1.879 & 3.691 & 0.018 & 0.105 & 2.015 & 1.998 & 1.236 & 0.835 & 0.681 & 0.127 & 1.238\\
PFF-gc075 & 1.803 & 1.566 & 4.477 & 0.098 & 0.213 & 3.037 & 2.367 & 1.930 & 1.080 & 0.483 & 0.273 & 1.307\\
PFF-gc079 & 1.477 & 2.431 & 3.056 & 0.094 & 0.202 & 1.978 & 1.926 & 1.823 & 1.102 & 0.395 & 0.138 & 5.134\\
PFF-gc081 & -0.334 & 2.255 & 1.857 & 0.021 & 0.084 & 1.420 & 2.015 & 1.330 & 0.303 & 0.505 & 0.169 & 0.217\\
PFF-gc082 & 0.805 & 1.851 & 4.602 & 0.037 & 0.150 & 2.936 & 2.159 & 1.861 & 1.734 & 0.634 & 0.644 & 1.479\\
\hline
\end{tabular}
\medskip
\end{table*}

\begin{table*}
\caption{Lick index errors measured for NGC~5128 globular clusters.}
\label{tab:LickErrors}
\begin{tabular}{@{}lccccccccccc}
\hline
ID & H$\delta_{\rm A}$ & H$\delta_{\rm F}$ & CN$_1$ & CN$_2$ &  Ca4227 & G4300 & H$\gamma_{\rm A}$ & H$\gamma_{\rm F}$ & Fe4383 & Ca4455 & Fe4531\\ 
   & (\AA) & (\AA) &  (mag) &  (mag) &  (\AA) &  (\AA) &  (\AA) &  (\AA) &  (\AA) &  (\AA) &  (\AA) \\
\hline
AAT101906 & 0.966 & 0.563 & 0.032 & 0.033 & 0.536 & 0.873 & 1.021 & 0.461 & 1.213 & 0.505 & 0.686\\
AAT103195 & 1.492 & 0.956 & 0.042 & 0.045 & 0.713 & 1.172 & 1.429 & 0.833 & 1.771 & 0.807 & 1.172\\
AAT106880 & 1.305 & 0.856 & 0.041 & 0.044 & 0.723 & 1.252 & 1.304 & 0.674 & 1.814 & 0.816 & 1.280\\
AAT107060 & 1.170 & 0.736 & 0.037 & 0.039 & 0.626 & 1.045 & 1.173 & 0.582 & 1.462 & 0.663 & 0.930\\
AAT107977 & 1.015 & 0.612 & 0.034 & 0.035 & 0.583 & 0.988 & 1.078 & 0.520 & 1.425 & 0.633 & 0.882\\
AAT109380 & 0.842 & 0.462 & 0.029 & 0.029 & 0.484 & 0.780 & 0.916 & 0.368 & 1.099 & 0.437 & 0.590\\
\end{tabular}
\end{table*}
\begin{table*}
\contcaption{}
\begin{tabular}{@{}lccccccccccc}
\hline
ID & H$\delta_{\rm A}$ & H$\delta_{\rm F}$ & CN$_1$ & CN$_2$ &  Ca4227 & G4300 & H$\gamma_{\rm A}$ & H$\gamma_{\rm F}$ & Fe4383 & Ca4455 & Fe4531\\ 
   & (\AA) & (\AA) &  (mag) &  (mag) &  (\AA) &  (\AA) &  (\AA) &  (\AA) &  (\AA) &  (\AA) &  (\AA) \\
\hline
HGHH-G219 & 1.023 & 0.599 & 0.033 & 0.033 & 0.546 & 0.883 & 1.033 & 0.466 & 1.232 & 0.532 & 0.714\\
HGHH-G251 & 1.098 & 0.655 & 0.035 & 0.037 & 0.610 & 0.966 & 1.143 & 0.558 & 1.358 & 0.607 & 0.825\\
HGHH-G277 & 1.000 & 0.581 & 0.034 & 0.035 & 0.579 & 0.974 & 1.076 & 0.514 & 1.384 & 0.606 & 0.881\\
HGHH-G279 & 0.998 & 0.584 & 0.035 & 0.036 & 0.607 & 1.091 & 1.035 & 0.457 & 1.473 & 0.660 & 0.956\\
HGHH-G293 & 0.970 & 0.582 & 0.033 & 0.035 & 0.568 & 0.950 & 1.053 & 0.493 & 1.365 & 0.593 & 0.845\\
HGHH-G342 & 1.095 & 0.638 & 0.035 & 0.037 & 0.592 & 0.978 & 1.156 & 0.562 & 1.390 & 0.637 & 0.882\\
\end{tabular}
\end{table*}
\begin{table*}
\contcaption{}
\begin{tabular}{@{}lccccccccccc}
\hline
ID & H$\delta_{\rm A}$ & H$\delta_{\rm F}$ & CN$_1$ & CN$_2$ &  Ca4227 & G4300 & H$\gamma_{\rm A}$ & H$\gamma_{\rm F}$ & Fe4383 & Ca4455 & Fe4531\\ 
   & (\AA) & (\AA) &  (mag) &  (mag) &  (\AA) &  (\AA) &  (\AA) &  (\AA) &  (\AA) &  (\AA) &  (\AA) \\
\hline
PFF-gc069 & 1.248 & 0.734 & 0.038 & 0.041 & 0.635 & 1.013 & 1.213 & 0.615 & 1.488 & 0.650 & 0.924\\
PFF-gc074 & 1.078 & 0.630 & 0.036 & 0.038 & 0.607 & 1.018 & 1.138 & 0.539 & 1.442 & 0.644 & 0.906\\
PFF-gc075 & 1.088 & 0.653 & 0.035 & 0.036 & 0.582 & 0.957 & 1.122 & 0.537 & 1.339 & 0.604 & 0.807\\
PFF-gc079 & 0.981 & 0.559 & 0.033 & 0.033 & 0.549 & 0.895 & 1.050 & 0.487 & 1.311 & 0.578 & 0.803\\
PFF-gc081 & 1.086 & 0.653 & 0.036 & 0.038 & 0.617 & 1.045 & 1.118 & 0.545 & 1.498 & 0.664 & 0.942\\
PFF-gc082 & 1.090 & 0.662 & 0.034 & 0.035 & 0.564 & 0.939 & 1.068 & 0.507 & 1.309 & 0.547 & 0.740\\
\end{tabular}
\end{table*}
\begin{table*}
\contcaption{}
\begin{tabular}{@{}lcccccccccccc}
\hline
ID & C$_2$4668 & H$\beta$ & Fe5015 & Mg$_1$ & Mg$_2$ & Mg $b$ & Fe5270 & Fe5335 & Fe5406 & Fe5709 & Fe5782 & Na D\\
   & (\AA) & (\AA) & (\AA) &  (mag) &  (mag) &  (\AA) &  (\AA) &  (\AA) &  (\AA) &  (\AA) &  (\AA) & (\AA) \\
\hline
AAT101906 & 1.058 & 0.412 & 0.731 & 0.015 & 0.020 & 0.398 & 0.383 & 0.447 & 0.325 & 0.318 & 0.251 & 0.382\\
AAT103195 & 1.746 & 0.668 & 1.329 & 0.018 & 0.023 & 0.627 & 0.659 & 0.768 & 0.556 & 0.473 & 0.421 & 0.573\\
AAT106880 & 1.915 & 0.690 & 1.479 & 0.019 & 0.025 & 0.722 & 0.786 & 0.896 & 0.643 & 0.536 & 0.489 & 0.685\\
AAT107060 & 1.359 & 0.516 & 1.020 & 0.016 & 0.022 & 0.512 & 0.544 & 0.635 & 0.459 & 0.406 & 0.359 & 0.460\\
AAT107977 & 1.361 & 0.510 & 1.043 & 0.016 & 0.022 & 0.531 & 0.554 & 0.637 & 0.465 & 0.430 & 0.371 & 0.510\\
AAT109380 & 0.923 & 0.359 & 0.614 & 0.014 & 0.019 & 0.361 & 0.344 & 0.409 & 0.299 & 0.301 & 0.236 & 0.362\\
\end{tabular}
\end{table*}
\begin{table*}
\contcaption{}
\begin{tabular}{@{}lcccccccccccc}
\hline
ID & C$_2$4668 & H$\beta$ & Fe5015 & Mg$_1$ & Mg$_2$ & Mg $b$ & Fe5270 & Fe5335 & Fe5406 & Fe5709 & Fe5782 & Na D\\
   & (\AA) & (\AA) & (\AA) &  (mag) &  (mag) &  (\AA) &  (\AA) &  (\AA) &  (\AA) &  (\AA) &  (\AA) & (\AA) \\
\hline
HGHH-G204 & 1.488 & 0.546 & 1.133 & 0.017 & 0.022 & 0.575 & 0.606 & 0.686 & 0.501 & 0.432 & 0.381 & 0.518\\
HGHH-G219 & 1.105 & 0.434 & 0.759 & 0.015 & 0.020 & 0.406 & 0.401 & 0.464 & 0.346 & 0.337 & 0.274 & 0.391\\
HGHH-G251 & 1.225 & 0.475 & 0.893 & 0.016 & 0.021 & 0.455 & 0.472 & 0.534 & 0.392 & 0.360 & 0.298 & 0.421\\
HGHH-G277 & 1.330 & 0.498 & 0.962 & 0.016 & 0.021 & 0.506 & 0.508 & 0.594 & 0.437 & 0.391 & 0.333 & 0.501\\
HGHH-G279 & 1.451 & 0.536 & 1.140 & 0.017 & 0.022 & 0.596 & 0.623 & 0.732 & 0.528 & 0.473 & 0.421 & 0.551\\
HGHH-G293 & 1.262 & 0.487 & 0.909 & 0.016 & 0.021 & 0.479 & 0.487 & 0.564 & 0.417 & 0.374 & 0.311 & 0.471\\
\end{tabular}
\end{table*}
\begin{table*}
\contcaption{}
\begin{tabular}{@{}lcccccccccccc}
\hline
ID & C$_2$4668 & H$\beta$ & Fe5015 & Mg$_1$ & Mg$_2$ & Mg $b$ & Fe5270 & Fe5335 & Fe5406 & Fe5709 & Fe5782 & Na D\\
   & (\AA) & (\AA) & (\AA) &  (mag) &  (mag) &  (\AA) &  (\AA) &  (\AA) &  (\AA) &  (\AA) &  (\AA) & (\AA) \\
\hline
PFF-gc069 & 1.370 & 0.541 & 1.007 & 0.016 & 0.022 & 0.516 & 0.524 & 0.598 & 0.453 & 0.390 & 0.332 & 0.452\\
PFF-gc074 & 1.335 & 0.504 & 0.960 & 0.016 & 0.021 & 0.495 & 0.503 & 0.589 & 0.433 & 0.383 & 0.327 & 0.464\\
PFF-gc075 & 1.268 & 0.488 & 0.922 & 0.016 & 0.021 & 0.478 & 0.480 & 0.555 & 0.405 & 0.369 & 0.308 & 0.444\\
PFF-gc079 & 1.184 & 0.436 & 0.842 & 0.015 & 0.020 & 0.463 & 0.460 & 0.534 & 0.389 & 0.358 & 0.302 & 0.417\\
PFF-gc081 & 1.407 & 0.524 & 1.054 & 0.016 & 0.022 & 0.533 & 0.546 & 0.639 & 0.475 & 0.410 & 0.353 & 0.502\\
PFF-gc082 & 1.148 & 0.448 & 0.824 & 0.015 & 0.020 & 0.431 & 0.426 & 0.496 & 0.361 & 0.348 & 0.284 & 0.413\\
\hline
\end{tabular}
\medskip
\end{table*}

\begin{table*}
\caption{Polynomial coefficients for Lick index--metallicity relations of the
Schiavon \etal(2005) Milky Way GCs from least-squares fitting. 
For a given index I, [M/H]$_{I}=a0+a1\times $I$+a2\times $I$^2$.
$r$ is the correlation coefficient of the fit and the rms is $\sqrt{\chi^2/n}$.}
\label{tab:polynomials}
\begin{tabular}{@{}llllll}
\hline
Index	&	{\it a0}&	{\it a1}&	{\it a2}	& {\it r} & 	rms \\
\hline
HdA	&	-0.6686	&	-0.2740	&	-0.0085	&	0.951	&	0.154	\\
HdF	&	-0.1620	&	-0.5950	&	0.0185	&	0.936	&	0.177	\\
CN$_1$	&	-0.9923	&	10.4795	&	-25.9525 &	0.939	&	0.173	\\
CN$_2$	&	-0.9397	&	9.9741	&	-26.3867 &	0.945	&	0.164	\\
Ca4227	&	-3.1711	&	4.5438	&	-1.5422	&	0.956	&	0.147	\\
G4300	&	-2.5675	&	0.6073	&	-0.0365	&	0.935	&	0.178	\\
H$\gamma_{\rm A}$	&	-0.6686	&	-0.2740	&	-0.0085	&	0.951	&	0.154	\\
H$\gamma_{\rm F}$	&	-0.1620	&	-0.5950	&	0.0185	&	0.936	&	0.177	\\
Fe4383	&	-2.2373	&	0.7071	&	-0.0539	&	0.959	&	0.142	\\
Ca4455	&	-6.8786	&	6.0187	&	-1.2593	&	0.929	&	0.185	\\
Fe4531	&	-3.4927	&	1.6508	&	-0.2138	&	0.908	&	0.210	\\
C$_2$4668	&	-1.1142	&	0.5493	&	-0.0749	&	0.927	&	0.188	\\
H$\beta$	&	-0.4124	&	0.4997	&	-0.3862	&	0.878	&	0.240	\\
Fe5015	&	---	&	---	&	---	&	---	&	---	\\
Mg$_1$	&	-2.2199	&	33.2872	&	-26.1750 &	0.917	&	0.199	\\
Mg$_2$	&	-2.4490	&	14.5595	&	-21.0308 &	0.954	&	0.150	\\
Mg $b$	&	-2.4214	&	0.8438	&	-0.0713	&	0.952	&	0.153	\\
Fe5270	&	-2.6124	&	1.3299	&	-0.1449	&	0.955	&	0.149	\\
Fe5335	&	-2.5049	&	1.4480	&	-0.1690	&	0.946	&	0.162	\\
Fe5406	&	-2.2713	&	1.9360	&	-0.3772	&	0.946	&	0.162	\\
Fe5709	&	-2.2739	&	2.2509	&	0.0105	&	0.921	&	0.195	\\
Fe5782	&	-2.1993	&	2.5623	&	-0.5959	&	0.821	&	0.286	\\
Na D	&	---	&	---	&	---	&	---	&	---	\\
\hline
\end{tabular}

\medskip
\end{table*}


\begin{thebibliography}{99}

\bibitem[Ashman \& Zepf(1992)]{1992ApJ...384...50A} Ashman, K.~M., \& Zepf, S.~E.\ 1992, ApJ, 384, 50 
\bibitem[Ashman et al.(1994)]{1994AJ....108.2348A} Ashman, K.~M., Bird, C.~M., 
\& Zepf, S.~E.\ 1994, AJ, 108, 2348 
\bibitem[Beasley et al.(2000)]{2000MNRAS.318.1249B} Beasley, M.~A., 
Sharples, R.~M., Bridges, T.~J., Hanes, D.~A., Zepf, S.~E., Ashman, K.~M., 
\& Geisler, D.\ 2000, MNRAS, 318, 1249 
\bibitem[Beasley et al.(2002)]{2002MNRAS.336..168B} Beasley, M.~A., Hoyle, 
F., \& Sharples, R.~M.\ 2002a, MNRAS, 336, 168 
\bibitem[Beasley et al.(2002)]{2002MNRAS.333..383B} Beasley, M.~A., Baugh, C.~M., Forbes, D.~A., 
Sharples, R.~M., \& Frenk, C.~S.\ 2002b, MNRAS, 333, 383 
\bibitem[Beasley et al.(2003)]{2003MNRAS.340..341B} Beasley, M.~A., Harris, 
W.~E., Harris, G.~L.~H., \& Forbes, D.~A.\ 2003, MNRAS, 340, 341 
\bibitem[Bekki et al.(2002)]{2002MNRAS.335.1176B} Bekki, K., Forbes, D.~A., Beasley, M.~A., \& Couch, W.~J.\ 2002, MNRAS, 335, 1176 
\bibitem[Bekki et al.(2003)]{2003MNRAS.338..587B} Bekki, K., Harris, W.~E., \& Harris, G.~L.~H.\ 2003, MNRAS, 338, 587 
\bibitem[Bekki \& Peng(2006)]{2006MNRAS.370.1737B} Bekki, K., \& Peng, E.~W.\ 2006, MNRAS, 370, 1737
\bibitem[Brodie \& Strader(2006)]{2006ARA&A..44..193B} Brodie, J.~P., \& Strader, J.\ 2006, 
ARA\&A, 44, 193
\bibitem[Burstein et al.(1984)]{1984ApJ...287..586B} Burstein, D., Faber, 
S.~M., Gaskell, C.~M., \& Krumm, N.\ 1984, ApJ, 287, 586 
\bibitem[Cenarro et al.(2007)]{2007AJ....134..391C} Cenarro, A.~J., Beasley, M.~A., Strader, J., 
Brodie, J.~P., \& Forbes, D.~A.\ 2007, AJ, 134, 391 
\bibitem[Clark(1975)]{1975ApJ...199L.143C} Clark, G.~W.\ 1975, ApJL, 199, L143 
\bibitem[Cohen et al.(1998)]{1998ApJ...496..808C} Cohen, J.~G., Blakeslee, J.~P., 
\& Ryzhov, A.\ 1998, ApJ, 496, 808 
\bibitem[Cohen et al.(2003)]{2003ApJ...592..866C} Cohen, J.~G., Blakeslee, J.~P., \& C{\^o}t{\'e}, P.\ 2003, ApJ, 592, 866
\bibitem[Cole et al.(2000)]{2000MNRAS.319..168C} Cole, S., Lacey, C.~G., Baugh, C.~M., 
\& Frenk, C.~S.\ 2000, MNRAS, 319, 168 
\bibitem[Colless et al.(2001)]{2001MNRAS.328.1039C} Colless, M., et al.\ 2001, MNRAS, 328, 1039 
\bibitem[Cote et al.(1998)]{1998ApJ...501..554C} C{\^o}t{\'e}, P., Marzke, R.~O., \& West, M.~J.\ 1998, ApJ, 501, 554 
\bibitem[De Angeli et al.(2005)]{2005AJ....130..116D} De Angeli, F., Piotto, G., Cassisi, S., Busso, G., 
Recio-Blanco, A., Salaris, M., Aparicio, A., \& Rosenberg, A.\ 2005, AJ, 130, 116 
\bibitem[Dufour et al.(1979)]{1979AJ.....84..284D} Dufour, R.~J., Harvel, 
C.~A., Martins, D.~M., Schiffer, F.~H., III, Talent, D.~L., Wells, D.~C., 
van den Bergh, S., \& Talbot, R.~J., Jr.\ 1979, AJ, 84, 284 
\bibitem[Forbes et al.(1997)]{1997AJ....113.1652F} Forbes, D.~A., Brodie, J.~P., \& Grillmair, C.~J.\ 1997, AJ, 113, 1652 
\bibitem[Harris(1996)]{1996AJ....112.1487H} Harris, W.~E.\ 1996, AJ, 112, 1487
\bibitem[Harris et al.(2006)]{2006ApJ...636...90H} Harris, W.~E., Whitmore, 
B.~C., Karakla, D., Oko{\'n}, W., Baum, W.~A., Hanes, D.~A., \& Kavelaars, J.~J.\ 2006a, ApJ, 636, 90 
\bibitem[Harris et al.(2006)]{2006AJ....132.2187H} Harris, W.~E., Harris, G.~L.~H., Barmby, P., 
McLaughlin, D.~E., \& Forbes, D.~A.\ 2006b, AJ, 132, 2187 
\bibitem[Harris et al.(2004)]{2004AJ....128..712H} Harris, G.~L.~H., et 
al.\ 2004, AJ, 128, 712 
\bibitem[Harris \& Harris(2002)]{2002AJ....123.3108H} Harris, W.~E., \& Harris, G.~L.~H.\ 2002, AJ, 123, 3108 
\bibitem[Harris \& Harris(2000)]{2000AJ....120.2423H} Harris, G.~L.~H., \& Harris, W.~E.\ 2000, AJ, 120, 2423 
\bibitem[Held et al.(2002)]{2002IAUS..207..269H} Held, E.~V., Federici, L., 
Cacciari, C., \& Testa, V.\ 2002, Extragalactic Star Clusters, IAU symposium 207, p269 
\bibitem[Hempel et al.(2007)]{2007A&A...463..493H} Hempel, M., 
Kissler-Patig, M., Puzia, T.~H., \& Hilker, M.\ 2007, A\&A, 463, 493 
\bibitem[Hesser et al.(1986)]{1986ApJ...303L..51H} Hesser, J.~E., Harris, 
H.~C., \& Harris, G.~L.~H.\ 1986, ApJL, 303, L51 
\bibitem[Houdashelt et al.(2002)]{2002AAS...201.1405H} Houdashelt, M.~L., Trager, S.~C., Worthey, G., \& Bell, 
R.~A.\ 2002, BAAS, 34, 1118 (H02)
\bibitem[Jord{\'a}n et al.(2002)]{2002ApJ...576L.113J} Jord{\'a}n, A., 
C{\^o}t{\'e}, P., West, M.~J., \& Marzke, R.~O.\ 2002, ApJL, 576, L113 
\bibitem[Kaviraj et al.(2005)]{2005A&A...439..913K} Kaviraj, S., Ferreras, I., Yoon, S.-J., \& 
Yi, S.~K.\ 2005, A\&A, 439, 913 
\bibitem[Kundu et al.(2005)]{2005ApJ...634L..41K} Kundu, A., et al.\ 2005, ApJL, 634, L41
\bibitem[Kundu \& Zepf(2007)]{2007ApJ...660L.109K} Kundu, A., \& Zepf, S.~E.\ 2007, ApJL, 660, L109
\bibitem[Larsen et al.(2001)]{2001AJ....1sta	21.2974L} Larsen, S.~S., Brodie, 
J.~P., Huchra, J.~P., Forbes, D.~A., \& Grillmair, C.~J.\ 2001, AJ, 121, 2974 
\bibitem[Larsen et al.(2005)]{2005A&A...443..413L} Larsen, S.~S., Brodie, 
J.~P., \& Strader, J.\ 2005, A\&A, 443, 413 
\bibitem[Lee et al.(2000)]{2000AJ....120..998L} Lee, H.-c., Yoon, S.-J., \& Lee, Y.-W.\ 2000, AJ, 120, 998
\bibitem[Lee \& Worthey(2005)]{2005ApJS..160..176L} Lee, H.-c., \& Worthey, G.\ 2005, ApJS, 160, 176 (LW05)
\bibitem[Lee et al.(2006)]{2006astro.ph..5425L} Lee, H.-c., Worthey, G., 
Trager, S.~C., \& Faber, S.~M.\ 2006, (astro-ph/0605425)
\bibitem[Lewis et al.(2002)]{2002MNRAS.333..279L} Lewis, I.~J., et al.\ 2002, MNRAS, 333, 279 
\bibitem[Malin(1978)]{1978Natur.276..591M} Malin, D.~F.\ 1978, Nature, 276, 591 
\bibitem[Maraston(1998)]{1998MNRAS.300..872M} Maraston, C.\ 1998, MNRAS, 300, 872 
\bibitem[Mendel et al.(2007)]{2007MNRAS.379.1618M} Mendel, J.~T., Proctor, R.~N., \& Forbes, D.~A.\ 2007, 
MNRAS, 379, 1618
\bibitem[Mould \& Aaronson(1980)]{1980ApJ...240..464M} Mould, J., \& 
Aaronson, M.\ 1980, ApJ, 240, 464 
\bibitem[Peng et al.(2002)]{2002AJ....124.3144P} Peng, E.~W., Ford, H.~C., 
Freeman, K.~C., \& White, R.~L.\ 2002, AJ, 124, 3144 
\bibitem[Peng et al.(2004)]{2004ApJ...602..705P} Peng, E.~W., Ford, H.~C., 
\& Freeman, K.~C.\ 2004, ApJ, 602, 705 (PFF04)
\bibitem[Peng et al.(2004)]{2004ApJS..150..367P} Peng, E.~W., Ford, H.~C., 
\& Freeman, K.~C.\ 2004b, ApJS, 150, 367 
\bibitem[Peng et al.(2004)]{2004ApJ...602..685P} Peng, E.~W., Ford, H.~C., 
\& Freeman, K.~C.\ 2004c, ApJ, 602, 685 
\bibitem[Peng et al.(2006)]{2006ApJ...639...95P} Peng, E.~W., et al.\ 2006, ApJ, 639, 95 
\bibitem[Perrett et al.(2002)]{2002AJ....123.2490P} Perrett, K.~M., Bridges, T.~J., 
Hanes, D.~A., Irwin, M.~J., Brodie, J.~P., Carter, D., Huchra, J.~P., \& Watson, F.~G.\ 2002, AJ, 123, 2490 
\bibitem[Pierce et al.(2006)]{2006MNRAS.366.1253P} Pierce, M., et al.\ 2006, MNRAS, 366, 1253 
\bibitem[Pipino et al.(2007)]{2007ApJ...665..295P} Pipino, A., Puzia, T.~H., \& Matteucci, F.\ 2007, ApJ, 665, 295
\bibitem[Pritzl et al.(2005)]{2005AJ....130.2140P} Pritzl, B.~J., Venn, K.~A., \& Irwin, M.\ 2005, AJ, 130, 2140 
\bibitem[Proctor et al.(2004)]{2004MNRAS.355.1327P} Proctor, R.~N., Forbes, D.~A., \& Beasley, M.~A.\ 2004, 
MNRAS, 355, 1327 
\bibitem[Puzia et al.(1999)]{1999AJ....118.2734P} Puzia, T.~H., Kissler-Patig, M., Brodie, J.~P., \& Huchra, J.~P.\ 1999, 
AJ, 118, 2734
\bibitem[Puzia et al.(2002)]{2002A&A...391..453P} Puzia, T.~H., Zepf, 
S.~E., Kissler-Patig, M., Hilker, M., Minniti, D., \& Goudfrooij, P.\ 2002a, A\&A, 391, 453
\bibitem[Puzia et al.(2002)]{2002A&A...395...45P} Puzia, T.~H., Saglia, R.~P., Kissler-Patig, M., 
Maraston, C., Greggio, L., Renzini, A., \& Ortolani, S.\ 2002b, A\&A, 395, 45 
\bibitem[Puzia et al.(2005)]{2005A&A...439..997P} Puzia, T.~H., 
Kissler-Patig, M., Thomas, D., Maraston, C., Saglia, R.~P., Bender, R., 
Goudfrooij, P., \& Hempel, M.\ 2005, A\&A, 439, 997 
\bibitem[Rejkuba(2004)]{2004A&A...413..903R} Rejkuba, M.\ 2004, A\&A, 413, 903 
\bibitem[Rejkuba et al.(2005)]{2005ApJ...631..262R} Rejkuba, M., Greggio, L., Harris, W.~E., Harris, G.~L.~H., 
\& Peng, E.~W.\ 2005, APJ, 631, 262 
\bibitem[Richardson \& Green (1997)] {} Richardson, S., \& Green, P.G., 1997, JR Stat. Soc. B, 59, 731
\bibitem[Salaris \& Cassisi(2007)]{2007A&A...461..493S} Salaris, M., \& Cassisi, S.\ 2007, A\&A, 461, 493 
\bibitem[Schiavon et al.(2005)]{2005ApJS..160..163S} Schiavon, R.~P., Rose, J.~A., 
Courteau, S., \& MacArthur, L.~A.\ 2005, ApJS, 160, 163 
\bibitem[Sharples(1988)]{1988IAUS..126..545S} Sharples, R.\ 1988, The Harlow-Shapley Symposium on Globular 
Cluster Systems in Galaxies, IAU symposium 126, p. 545
\bibitem[Sheather \& Jones (1991)]{}Sheather, S.J., \& Jones, M.C. \ 1991, J R Stat. Soc. B, 53, 683
\bibitem[Silverman(1986)]{1986desd.book.....S} Silverman, B.~W.\ 1986, Monographs on Statistics and 
Applied Probability, London: Chapman and Hall, 1986  
\bibitem[Strader et al.(2004)]{2004AJ....127.3431S} Strader, J., Brodie, 
J.~P., \& Forbes, D.~A.\ 2004, AJ, 127, 3431 
\bibitem[Strader et al.(2005)]{2005AJ....130.1315S} Strader, J., Brodie, 
J.~P., Cenarro, A.~J., Beasley, M.~A., \& Forbes, D.~A.\ 2005, AJ, 130, 1315 
\bibitem[Strader et al.(2006)]{2006AJ....132.2333S} Strader, J., Brodie, J.~P., Spitler, L., \& 
Beasley, M.~A.\ 2006, AJ, 132, 2333 
\bibitem[Strader et al.(2007)]{2007AJ....133.2015S} Strader, J., Beasley, M.~A., \& Brodie, J.~P.\ 2007, AJ, 133, 2015 
\bibitem[Taylor et al.(1996)]{1996ASPC..101..195T} Taylor, K., Bailey, J., Wilkins, T., Shortridge, 
K., \& Glazebrook, K.\ 1996, Astronomical Data Analysis Software and Systems V, 
George H. Jacoby and Jeannette Barnes, eds. Vol 101, p 195 
\bibitem[Thomas et al.(2003)]{2003MNRAS.339..897T} Thomas, D., Maraston, C., \& Bender, R.\ 2003, MNRAS, 339, 897
\bibitem[Thomas et al.(2004)]{2004MNRAS.351L..19T} Thomas, D., Maraston, C., \& Korn, A.\ 2004, MNRAS, 351, L19 (TMK04)
\bibitem[Tonry \& Davis(1979)]{1979AJ.....84.1511T} Tonry, J., \& Davis, M.\ 1979, AJ, 84, 1511 
\bibitem[Trager et al.(1998)]{1998ApJS..116....1T} Trager, S.~C., Worthey, G., Faber, S.~M., 
Burstein, D., \& Gonzalez, J.~J.\ 1998, ApJS, 116, 1 
\bibitem[van den Bergh et al.(1981)]{1981AJ.....86...24V} van den Bergh, S., Hesser, J.~E., 
\& Harris, G.~L.~H.\ 1981, AJ, 86, 24 
\bibitem[VanDalfsen \& Harris(2004)]{2004AJ....127..368V} VanDalfsen, M.~L., \& Harris, W.~E.\ 2004, AJ, 127, 368 
\bibitem[Vazdekis(1999)]{1999ApJ...513..224V} Vazdekis, A.\ 1999, ApJ, 513, 224 
\bibitem[Woodley et al.(2005)]{2005AJ....129.2654W} Woodley, K.~A., Harris, W.~E., \& 
Harris, G.~L.~H.\ 2005, AJ, 129, 2654 
\bibitem[Woodley et al.(2007)]{2007AJ....134..494W} Woodley, K.~A., Harris, W.~E., Beasley, M.~A., 
Peng, E.~W., Bridges, T.~J., Forbes, D.~A., \& Harris, G.~L.~H.\ 2007, AJ, 134, 494 
\bibitem[Worthey \& Ottaviani(1997)]{1997ApJS..111..377W} Worthey, G., \& Ottaviani, D.~L.\ 1997, ApJS, 111, 377 
\bibitem[Yoon et al.(2006)]{2006Sci...311.1129Y} Yoon, S.-J., Yi, S.~K., \& Lee, Y.-W.\ 2006, Science, 311, 1129 
\bibitem[Zinn \& West(1984)]{1984ApJS...55...45Z} Zinn, R., \& West, M.~J.\ 1984, ApJS, 55, 45
\end{thebibliography}
\end{document}